\documentstyle[12pt]{article}

\def\thebibliography#1{%
\parindent 0em
\vspace{9pt}
\begin{flushleft}\large\bf {References}\end{flushleft}
\addvspace{3pt}\nopagebreak\list
{[\arabic{enumi}]} {\settowidth\labelwidth{[#1]}
\leftmargin\labelwidth
\leftmargin=17pt
 \advance\leftmargin\labelsep
 \usecounter{enumi}\@bibsetup}
\def\newblock{\hskip .11em plus .33em minus -.07em}
 \sloppy\clubpenalty4000\widowpenalty4000
 \sfcode`\.=1000\relax}

\def\and{
{\LARGE\em \&}
\bigskip
}

\def\@bibsetup{
\itemsep=0pt \parsep=0pt
\small}

\newtheorem{theorem}{Theorem}
	
	\newtheorem{lemma}{Lemma}
	
	\newtheorem{definition}{Definition}
\def\Cyl{{\rm Cyl}}
\def\LG{{\cal L}_G}
\def\Ad{{\rm Ad}}
\def\be{\begin{equation}}
\def\ee{\end{equation}}
\def\ba{\begin{eqnarray}}
\def\ea{\end{eqnarray}}
\def\Gra{{\rm Gra}}
\def\Gras{{\rm Gra}(\Sigma)}
\def\R{{\cal R}}
\def\E{{\cal E}}
\def\A{{\cal A}}
\def\G{{\cal G}}
\def\ag{{{\cal A}/{\cal G}}}

\def\agb{{\overline {{\cal A}/{\cal G}}}}

\def\S{\Sigma}\def\g{\gamma}\def\Gb{{\overline \G}}

\def\Ab{{\overline \A}}
\def\Gb{{\overline \G}}
\def\B{\agb}
\def\a{\alpha}

\def\g{\gamma}
\def\d{\delta}
\def\Comp{{\mathchoice
{\setbox0=\hbox{$\displaystyle\rm C$}\hbox{\hbox to0pt
{\kern0.4\wd0\vrule height0.9\ht0\hss}\box0}}
{\setbox0=\hbox{$\textstyle\rm C$}\hbox{\hbox to0pt
{\kern0.4\wd0\vrule height0.9\ht0\hss}\box0}}
{\setbox0=\hbox{$\scriptstyle\rm C$}\hbox{\hbox to0pt
{\kern0.4\wd0\vrule height0.9\ht0\hss}\box0}}
{\setbox0=\hbox{$\scriptscriptstyle\rm C$}\hbox{\hbox to0pt
{\kern0.4\wd0\vrule height0.9\ht0\hss}\box0}}}}
\def\Co{{\mathchoice
{\setbox0=\hbox{$\displaystyle\rm C$}\hbox{\hbox to0pt
{\kern0.4\wd0\vrule height0.9\ht0\hss}\box0}}
{\setbox0=\hbox{$\textstyle\rm C$}\hbox{\hbox to0pt
{\kern0.4\wd0\vrule height0.9\ht0\hss}\box0}}
{\setbox0=\hbox{$\scriptstyle\rm C$}\hbox{\hbox to0pt
{\kern0.4\wd0\vrule height0.9\ht0\hss}\box0}}
{\setbox0=\hbox{$\scriptscriptstyle\rm C$}\hbox{\hbox to0pt
{\kern0.4\wd0\vrule height0.9\ht0\hss}\box0}}}}
\def\Rl{{\mathchoice
{\setbox0=\hbox{$\displaystyle\rm R$}\hbox{\hbox to0pt
{\kern0.4\wd0\vrule height0.9\ht0\hss}\box0}}
{\setbox0=\hbox{$\textstyle\rm R$}\hbox{\hbox to0pt
{\kern0.4\wd0\vrule height0.9\ht0\hss}\box0}}
{\setbox0=\hbox{$\scriptstyle\rm R$}\hbox{\hbox to0pt
{\kern0.4\wd0\vrule height0.9\ht0\hss}\box0}}
{\setbox0=\hbox{$\scriptscriptstyle\rm R$}\hbox{\hbox to0pt
{\kern0.4\wd0\vrule height0.9\ht0\hss}\box0}}}}
\def\L{{\rm L}}

\def\Acb{\overline{{\A}^{\Co}}}
\def\Gcb{\overline{\G}^{\Co}}
\def\agbc{{\overline { {\cal A}^\Comp/{\cal G}^\Comp}}}

\def\o{\overline}
\def\ov{\overline}
\def\p{p}
\def\n{\nu}
\def\bs{\bigskip}

\def\wt{\widetilde}
\def\ov{\overline}

\def\wh{\widehat}

\def\agbc{{\overline {{\cal A}^\Co/{\cal G}^\Co}}}
\def\abc{{\overline {{\cal A}^\Co }}}
\def\gbc{{\overline {{\cal G}^\Co}}}

\begin{document}

\title{ Coherent State  Transforms for Spaces of Connections}

\author{Abhay Ashtekar\thanks{
 Center for Gravitational Physics and
Geometry, Physics Department, The Pennsylvania State University,
University Park,
PA 16802-6300, USA.}
{}~~~~~
Jerzy Lewandowski\thanks{Institute of Theoretical Physics,
University of Warsaw, 00-681 Warsaw, Poland}
{}~~~~~
Donald Marolf\thanks{Department of Physics, The University of California,
Santa Barbara, CA 93106, USA} \\
Jos\'e Mour\~ao
\thanks{Sector de F\'{\i}sica da U.C.E.H., Universidade do Algarve,
Campus de Gambelas, 8000 Faro, Portugal}
{}~~~~~
Thomas Thiemann$^\ast$}

\maketitle


\begin{abstract}

The Segal-Bargmann transform plays an important role in quantum
theories of linear fields. Recently, Hall obtained a non-linear analog
of this transform for quantum mechanics on Lie groups. Given a
compact, connected Lie group $G$ with its normalized Haar measure
$\mu_H$, the Hall transform is an isometric isomorphism from $L^2(G,
\mu_H)$ to ${\cal H}(G^{\Co})\cap L^2(G^{\Co}, \nu)$, where
$G^{\Co}$ the complexification of $G$, ${\cal H}(G^{\Co})$ the space
of holomorphic functions on $G^{\Co}$, and $\nu$ an appropriate
heat-kernel measure on $G^{\Co}$. We extend the Hall transform to the
infinite dimensional context of non-Abelian gauge theories by
replacing the Lie group $G$ by (a certain extension of) the space
${\cal A}/{\cal G}$ of connections modulo gauge transformations. The
resulting ``coherent state transform'' provides a holomorphic
representation of the holonomy $C^\star$ algebra of real gauge fields.
This representation is expected to play a key role in a
non-perturbative, canonical approach to quantum gravity in
4-dimensions.

\end{abstract}

\vfill \eject

{\Large {\bf Contents}}

\bigskip

\begin{itemize}

\item[{\bf  1.}] Introduction
\item[{\bf 2.}] Hall transform for compact groups $G$
\item[{\bf 3.}] Measures on spaces of connections

\begin{itemize}

\item[{\bf 3.1.}] Spaces $\Ab, \o {\cal G}$ and $\agb$
\item[{\bf 3.2.}] Measures on $\Ab$

\end{itemize}

\item[{\bf 4.}] Coherent state transforms for theories
of connections
\item[{\bf 5.}] Gauge covariant coherent
state transforms

\begin{itemize}

\item[{\bf 5.1.}] The transform and the main result
\item[{\bf 5.2.}] Consistency
\item[{\bf 5.3.}] Measures on $\Ab^\Co$
\item[{\bf 5.4.}] Gauge covariance

\end{itemize}

\item[{\bf 6.}] Gauge and diffeomorphism covariant
coherent state transforms

\begin{itemize}

\item[{\bf 6.1.}] The transform and the main  result
\item[{\bf 6.2.}] Consistency
\item[{\bf 6.3.}] Analyticity
\item[{\bf 6.4.}] Gauge and diffeomorphism covariance
\item[{\bf 6.5.}] Isometry

\end{itemize}


\item[] Appendix: The Abelian case

\end{itemize}
\vfill  \eject

\begin{section}{Introduction}

In the early sixties, Segal \cite{IES1,IES2} and Bargmann \cite{VB}
introduced an integral transform that led to a holomorphic
representation of quantum states of linear, Hermitian, Bose fields.
(For a review of the holomorphic --or, coherent-state--
representation, see Klauder \cite{KLAU}.)  The purpose of this paper
is to extend that construction to non-Abelian gauge fields and, in
particular, to general relativity. The key idea is to combine two
ingredients: i) A non-linear analog of the Segal-Bargmann transform
due to Hall \cite{Ha} for a system whose configuration space is a
compact, connected Lie group; and, ii) A calculus on the space of
connections modulo gauge transformations based on projective
techniques [6-15].

Let us begin with a brief summary of the overall situation. Recall
first that, in theories of connections, the classical configuration
space is given by $\ag$, where $\A$ is the space of connections on a
principal fibre bundle $P(\Sigma, G)$ over a (``spatial'') manifold
$\Sigma$, and $\G$ is the group of vertical automorphisms of $P$. In
this paper, we will assume that $\Sigma$ is an analytic n-manifold,
$G$ is a compact, connected Lie group, and elements of $\A$ and $\G$
are all smooth. In field theory the quantum configuration space is,
generically, a suitable completion of the classical one.  A candidate,
$\agb$, for such a completion of $\ag$ was recently introduced
\cite{AI}.  This space will play an important role throughout our
discussion. It first arose as the Gel'fand spectrum of a $C^\star$
algebra constructed from the so-called Wilson loop functions, the
traces of holonomies of smooth connections around (piecewise analytic)
closed loops. It is therefore a compact, Hausdorff space. However, it
was subsequently shown \cite{MM,AL3} that, using a suitable projective
family, $\agb$ can also be obtained as the projective limit of
topological spaces $G^n/{\rm Ad}$, the quotient of $G^n$ by the
adjoint action of $G$.  Here, we will work with this characterization
of $\agb$.

It turns out that $\agb$ is a very large space: there is a precise
sense in which it can be regarded as the ``universal home'' for
measures
\footnote{While we will be mostly concerned here with Hilbert spaces
of quantum states, the space $\agb$ is also useful in the Euclidean
approach to quantum gauge theories. In particular, the 2-dimensional
Yang-Mills theory can be constructed on $\R^2$ or on $S^1\times \R$ by
defining the appropriate measure on $\agb$ \cite{ALMMT1}.}
that define quantum quange theories in which the Wilson loop operators
are well-defined \cite{AMM}. However, it is small enough to admit
various notions from differential geometry such as forms, vector
fields, Laplacians and heat kernels \cite{AL2}.  In Yang-Mills
theories, one expects the physically relevant measures to have support
on a ``small'' subspace of $\agb$.  The structure of quantum general
relativity, on the other hand, is quite different. In the canonical
approach, each quantum state arises as a measure and there are strong
indications that measures with support on all of $\agb$ will be
physically significant \cite{AA2}.

Now, as in linear theories \cite{IES1}, for non-Abelian gauge fields,
it is natural to first construct a ``Schr\"odinger-type''
representation in which the Hilbert space of states arises as
$L^2(\agb , \mu)$ for a suitable measure $\mu$ on $\agb$. This will be
our point of departure. The projective techniques referred to above
enable us to define measures as well as integrals over $\agb$ as
projective limits of measures and integrals over $G^n/{\rm Ad}$.  We
would, however, like to construct a ``holomorphic representation''.
Thus, we need to complexify $\agb$, consider holomorphic functions
thereon and introduce suitable measures to integrate these functions.
It is here that we use the techniques introduced by Hall
\cite{Ha}.  Given any compact Lie group $G$, Hall considers its
complexification $G^{\Co}$, defines holomorphic functions on
$G^{\Co}$, and, using heat-kernel methods, introduces measures $\nu$
with appropriate fall-offs (for the scalar products between
holomorphic functions to be well-defined).  Finally, he provides a
transform $C_{\nu}$, from $L^2(G, \mu_H)$ to the space of
$\nu$-square-integrable holomorphic functions over $G^{\Co}$. Since
Hall's transform is of a geometric rather than algebraic or
representation-theoretic nature, it can be readily combined with the
projective techniques.  Using it, we will construct the appropriate
Hilbert spaces of holomorphic functions on $\overline{{\cal A}^{\Co}/
{\cal G}^{\Co}}$ --an appropriate complexification of $\agb$-- and
obtain isometric isomorphisms between this space and $L^2(\agb, \mu)$.
For gauge theories --such as the 2-dimensional Yang-Mills theory-- our
results provide a new, coherent state representation of quantum states
which is well suited to analyze a number of issues.

The main motivation for this analysis comes, however, from quantum
general relativity: the holomorphic representation serves as a key
step in the canonical approach to quantum gravity.  Let us make a
brief detour to explain this point. The canonical quantization program
for general relativity was initiated by P.A.M. Dirac and P. Bergmann
already in the late fifties, and developed further, over the next two
decades, by a number of researchers including R. Arnowitt, S.  Deser,
C. W. Misner and J. A. Wheeler and his co-workers.  The first step is
a reformulation of general relativity as a Hamiltonian system. This
was accomplished using 3-metrics as configuration variables rather
early. While these variables are natural from a geometrical point of
view, it turns out that they are not convenient for discussing the
dynamics of the theory.  In particular, the basic equations are {\it
non-polynomial} in these variables.  Therefore, a serious attempt at
making mathematical sense of their quantum analogs has never been made
and the work in this area has remained heuristic.

In the mid-eighties, however, it was realized \cite{AA1} that a
considerable simplification occurs if one uses self-dual connections
as dynamical variables. In particular, the basic equations become low
order polynomials. Furthermore, since the configuration variables are
now connections, one can take over the sophisticated machinery that
has been used to analyze gauge theories.  Consequently, over the last
few years, considerable progress could be made in this area. (For a
review, see, e.g., \cite{AA3}). However, in the Lorentzian signature,
self-dual connections are complex and provide a {\it complex}
coordinatization of the {\it phase space} of general relativity rather
than a real coordinatization of its configuration space.  Therefore,
if one is to base one's quantum theory on these variables, it is clear
heuristically that the quantum states must be represented by {\it
holomorphic} functionals of self-dual connections. (Detailed
considerations show that they should in fact be complex measures
rather than functionals.) Given the situation in the classical theory,
this is the representation in which one might expect the quantum
dynamics to simplify considerably. Indeed, heuristic treatments have
yielded a variety of results in support of this belief
\cite{RS,AA3}. Furthermore, they have brought out a potentially deep
connection between knot theory and quantum gravity \cite{Ba3}.  To
make these results precise, one first needs to construct the
holomorphic representation rigorously. The coherent state transform of
this paper provides a solution to this problem. In particular, it has
already led to a rigorous understanding of the relation between knots
and states of quantum gravity \cite{AA2,ALMMT2}.

The paper is organized as follows. In section 2, we recall the
definition and properties of the Hall transform. Section 3 summarizes
the relevant results from calculus on the space of connections. In
particular, in section 3, we will: i) construct, using projective
techniques, the spaces $\Ab$ of generalized connections, $\Gb$ of
generalized automorphisms of $P$ and their quotient $\agb$ and
complexifications ${\A}^{\Co}$ and ${\G}^{\Co}$; ii) see that the
space $\Ab$ is equipped with a natural measure $\mu_0$ which is
faithful and invariant under the induced action of the diffeomorphism
group of the underlying manifold $\Sigma$; and, iii) show that it also
admits a family of diffeomorphism invariant measures
$\mu^{(m)}$, introduced by Baez. All these measures project
down unambiguously to $\agb$. Section 4 contains a precise formulation
of the main problem of this paper and summary of our strategy.  In
section 5, using heat kernel methods, we construct a family of
(cylindrical) measures $\nu^{l}_t$ on $\Acb$, and a family of
transforms $Z^l_t$ from $L^2(\Ab, \mu_0)$ to (the Cauchy
completion of) the intersection ${\cal H}_{\cal C}\cap L^2(\Acb ,
\nu^l_t)$ of the space of cylindrical holomorphic functions on
$\Acb$ with the space of $\nu$-square integrable functions.
These transforms provide isometric isomorphisms between the two
spaces. Furthermore, the transforms are gauge-covariant so that they
map $\Gb$-invariant functions on $\Ab$ to $\Gcb$ invariant functions
on $\Acb$. However, these transforms are not diffeomorphism covariant:
Although the measure $\mu_0$ on $\Ab$ {\it is} diffeomorphism
invariant, to define the corresponding heat kernel one is forced to
introduce an additional structure which fails to be diffeomorphism
invariant \cite{AL2}. The Baez measures $\mu^{(m)}$, on the other
hand, are free of this difficulty. That is, using $\mu^{(m)}$ in
place of $\mu_0$, one can obtain coherent state transforms which are
both gauge {\it and} diffeomorphism covariant. This is the main result
of section 6. The Appendix provides the explicit expression of one of
these transforms for the case when the gauge group is Abelian.
\end{section}

\begin{section} {Hall transform for compact groups $G$}
\label{Hall}

In this section we recall from \cite{Ha} those aspects of the Hall
transform which will be needed in our main analysis.  Let $G^\Co$ be
the complexification of $G$ in the sense of \cite{Ho} and $\nu$ be a
bi-$G$-invariant measure on $G^\Co$ that falls off rapidly at infinity
(see (\ref{foff}) below).  The Hall transform $C_{\nu}$ is an
isometric isomorphism from $L^2(G, \mu_H)$, where $\mu_H$ denotes the
normalized Haar measure on $G$, onto the space of $\nu$-square
integrable holomorphic functions on $G^\Co$
\be
\label{HT0} C_{\nu} \ :  \ L^2(G, \mu_H) \
\rightarrow \ {\cal H} (G^\Co)
    \cap L^2(G^\Co, \nu(g^\Co)).
\ee
Such a transform exists whenever the Radon-Nikodym derivative
$d\nu/d\mu_H^{\Co}$ exists, is locally bounded away from zero, and
falls off at infinity in such a way that the integral
\be
\sigma_{\pi}^{\nu} = {1
\over dim V_{\pi}} \int_{G^\Co} \parallel
\pi (g^{\Co^{-1}}) \parallel^2 d\nu(g^\Co)
\label{foff}
\ee
is finite for all $\pi$.  Here, $\mu_H^{\Co}$ is the Haar measure on
$G^{\Co}$, $\pi$ denotes (one representative of) an isomorphism class
of irreducible representations of $G$ on the complex linear spaces
$V_{\pi}$, and, $|| A || = \sqrt{Tr (A^{\dagger}A)}$ for $A \in End
V_{\pi}$ and $A^{\dagger}$ the adjoint of $A$ with respect to a
$G$-invariant inner product on $V_{\pi}$.  For a $\nu$ satisfying
(\ref{foff}), the Hall transform is given by
\be
\label{HT}
\! [C_{\nu}(f)](g^\Co) = (f \star \rho_{\nu})(g^\Co) =
\int_{G} f(g)\  \rho_{\nu}(g^{-1}g^\Co)\ d\mu_H(g)   \ ,
\label{roce}
\ee
where $\rho_{\nu}(g^\Co)$ is the kernel of the transform given in
terms of $\nu$ by
\be
\label{U}
\rho_{\nu}(g^\Co) = \sum_{\pi} {dim V_{\pi} \over
 \sqrt{\sigma_{\pi}^{\nu}}}
Tr \big(\pi (g^{\Co^{-1}}) \big) \ .
\ee

The transform $C_{\nu}$ takes a particularly simple form for the (real
analytic) functions $k_{\pi ,A}$ on $G$ corresponding to matrix
elements of $\pi (g)$,
$$ k_{\pi ,A}(g) = Tr (\pi (g) A)\ .  $$
This is significant because, according to the Peter-Weyl Theorem the
matrix elements $k_{\pi,A}$, for all $\pi$ and all $A \in End
V_{\pi}$, span a dense subspace in $L^2(G, d\mu_H)$.  The image of
these functions $k_{\pi,A}$ under the transform is (see \cite{Ha})
\ba
 [C_{\nu}(k_{\pi,A})](g^\Co) &=&
[k_{\pi,A} \star \rho_{\nu}](g^\Co) \nonumber \\
&=& {1 \over \sqrt{\sigma_{\pi}^{\nu}}}  k_{\pi,A}(g^\Co ) \ .
\ea

The evaluation of the Hall transform of a generic function $f$, $f \in
L^2(G, d\mu_H)$, can be naturally divided into two steps.  In the
first, one obtains a real analytic function on the original group $G$,
$$ f \mapsto f\star \rho_{\nu} \ .  $$
In the second step the function $f \star \rho_{\nu}$ is analytically
continued to $G^\Co$. It follows from (\ref{U}) that
\be
\label{5b}
f \star \rho_{\nu} = \rho_{\nu} \star f.
\ee

A natural choice for the measure $\nu$ on $G^\Co$ is the ``averaged"
heat kernel measure $\nu_t$ \cite{Ha}.  This measure is defined by
\be
\label{avhk} d\nu_t (g^{\Co})
 = \Bigl[ \int_G \mu^\Co_t (g g^\Co) d\mu_H(g)\Bigr]
d\mu_H^{\Co}(g^\Co)\ ,
\ee
where $\mu^\Co_t$ is the heat kernel on $G^\Co$; i.e., the solution to
the equations
\ba
\label{mut} { \partial \over \partial t} \mu^\Co_t &=& {1
\over 4} \Delta_{G^\Co} \mu^\Co_t \nonumber \\
\mu^\Co_0(g^\Co) &=& \delta(g^\Co, 1_{G^\Co}) \ .
\ea
Here the Laplacian $\Delta_{G^{\Co}}$ is defined by a left
$G^{\Co}$-invariant, bi-$G$-invariant metric on $G^{\Co}$, $1_{G^\Co}$
denotes the identity of the group $G^\Co$, and $\d$ is the delta
function corresponding to the measure $\mu_H^\Co$. If we take for
$\nu$ the averaged heat kernel measure $\nu_t$ then in
(\ref{foff}) we have
\be
 \sigma^{\nu_t}_{\pi} = e^{t \d_{\pi}} \ , \label{ev}
\ee
where $\d_{\pi}$ denotes the eigenvalue of the Laplacian $\Delta_G$ on
$G$ corresponding to the eigenfunction $k_{\pi,A}$. Notice that
$\Delta_G$ gives the representation on $L^2(G, d\mu_H)$ of a (unique
up to a multiplicative constant if $G$ is simple) quadratic Casimir
element. The result (\ref{ev}) follows from (\ref{U}) and the fact
that the kernel $\rho_{\nu_t} \equiv \rho_t$ of the transform
$C_{\nu_t} \equiv C_t$ is the (analytic extension of) the
fundamental solution of the heat equation on $G$: \be {\partial \over
\partial t} \rho_t = {1 \over 2}
\Delta_G \rho_t \ .
\ee
Therefore, in this case one obtains
\be
\rho_t(g^\Co) = \sum_{\pi} dim V_{\pi}\; e^{- t \delta_{\pi}/2}
\; \mbox{Tr} (\pi (g^{\Co^{-1}})).
\ee
These results will be used in sections 4 and 5 to define infinite
dimensional generalizations of the Hall transform.
\end{section}

\begin{section}{Measures on spaces of connections}
\label{measures}

In this section, we will summarize the construction of certain spaces
of generalized connections and indicate how one can introduce
interesting measures on them. Since the reader may not be familiar
with any of these results, we will begin with a chronological sketch
of the development of these ideas.

Recall that, in field theories of connections, a basic object is the
space $\cal A$ of smooth connections on a given smooth principal fibre
bundle $P(\Sigma, G)$. (We will assume the base manifold $\Sigma$ to
be analytic and $G$ to be a compact, connected Lie group.)  The
classical configuration space is then the space $\ag$ of orbits in
$\cal A$ generated by the action of the group $\cal G$ of smooth
vertical automorphisms of $P$.  In quantum mechanics,
the domain space of quantum states coincides with the classical
configuration space. In quantum field theories, on the other hand, the
domain spaces are typically larger; indeed the classical configuration
spaces generally form a set of zero measure. In gauge theories,
therefore, one is led to the problem of finding suitable extensions of
$\ag$. The problem is somewhat involved because $\ag$ is a rather
complicated, {\it non-linear} space.

One avenue \cite{AI} towards the resolution of this problem is offered by
the the Gel'fand-Naimark theory of commutative $C^\star$-algebras.
Since traces of holo\-no\-mies of connections around closed loops are
gauge invariant, one can use them to construct a certain Abelian
$C^\star$-algebra with identity, called the {\it holonomy algebra}.
Elements of this algebra separate points of $\ag$, whence, $\ag$ is
densely embedded in the spectrum of the algebra. The spectrum is
therefore denoted by $\agb$. This extension of $\ag$ can be taken to
be the domain space of quantum states. Indeed, in every cyclic
representation of the holonomy algebra, states can be identified as
elements of $L^2(\agb ,\mu)$ for some regular Borel measure $\mu$ on
$\agb$.

One can characterize the space $\agb$ purely algebraically
\cite{AI,AL1} as the space of all homomorphisms from a certain
group (formed out of piecewise analytic, based loops in $\Sigma$) to
the structure group $G$.  Another --and, for the present paper more
convenient-- characterization can be given using certain projective
limit techniques \cite{MM,AL3}: $\agb$ with the Gel'fand topology is
homeomorphic to the projective limit, with Tychonov topology, of an
appropriate projective family of finite dimensional compact spaces.
This result simplifies the analysis of the structure of $\agb$
considerably.  Furthermore, it provides an extension of $\ag$ also in
the case when the structure group $G$ is {\it non-compact}.
Projective techniques were first used in \cite{MM,AL3} for
measure-theoretic purposes and then extended in \cite{AL2} to
introduce ``differential geometry'' on $\agb$.

The first example of a non-trivial measure on $\agb$ was constructed
in \cite{AL1} using the Haar measure on the structure group $G$. This
is a natural measure in that it does not require any additional input;
it is also faithful and invariant under the induced action of the
diffeomorphism group of $\Sigma$. Baez \cite{Ba1} then proved that
every measure on $\agb$ is given by a suitably consistent family of
measures on the projective family. He also replaced the projective
family labeled by loops on $\S$ \cite{MM,AL3} by a family labeled by
graphs (see also \cite{Ba2,L}) and introduced a family of measures
which depend on characteristics of vertices. Finally, he provided a
diffeomorphism invariant construction which, given a family of
preferred vertices and almost any measure on $G$, produces a
diffeomorphism invariant measure on $\agb$.

We will now provide the relevant details of these constructions.  Our
treatment will, however, differ slightly from that of the papers
cited above.

\begin{subsection} {Spaces $\o{\cal A}, \o{\cal G}$ and $\agb$}
\label{031}

Let $\Sigma$ be a connected analytic $n$-manifold and $G$ be a
compact, connected Lie group. Consider the set $\cal E$ of all
oriented, unparametrized, embedded, analytic intervals
(edges) in $\Sigma$. We introduce the space $\Ab$
of (generalized) connections on $\Sigma$ as the space of all maps
${\o A}\ :\ {\cal E}\ \rightarrow G$, such that
\be
\label{10}
\o A(e^{-1}) = \o [A(e)]^{-1} , \ \ \  {\rm and} \ \ \
{\o A}(e_2\circ e_1)\ =\ {\o A}(e_2){\o A} ( e_1)
\ee
whenever two edges $e_2,e_1\in{\cal E}$ meet to form an edge.  Here,
$e_2 \circ e_1$ denotes the standard path product and $e^{-1}$ denotes
$e$ with opposite orientation.  The group $\o\G$ of (generalized)
gauge transformations acting on $\Ab$ is the space of maps $\ov
g:\Sigma \rightarrow G$ or equivalently the Cartesian product group
\be
\label{11} \Gb\ := \times_{x\in\S} \ G.
\ee
A gauge transformation $\ov g \in \Gb$ acts on $\ov A \in \Ab$ through
\be
[\overline{g}(\overline{A})](e_{p_1,p_2}) =
\overline{g}_{p_1} \overline{A}(e_{p_1,p_2}) (\o{g}_{p_2})^{-1}
\ee
where $e_{p_1,p_2}$ is an edge from $p_1 \in \Sigma$ to $p_2 \in
\Sigma$ and $\o{g}_{p_i}$ is the group element assigned to $p_i$ by
$\o{g} $.  The space $\Gb$ equipped with the product topology is a
compact topological group.  Note also that $\Ab$ is a closed subset of
\be
\Ab\subset \times_{e\in\E} \ \A_e ,
\ee
where the space $\A_e$ of all maps from the one point set $\{e\}$ to $
G$ is homeomorphic to $G$.  $\Ab$ is then compact in the topology
induced from this product.

It turns out that the space $\Ab$ (and also $\Gb$) can be regarded as
the projective limit of a family labeled by graphs in $\Sigma$ in
which each member is homeomorphic to a finite product of copies of $G$
\cite{MM,AL3}. Since this fact will be important for describing measures
on $\Ab$ and for constructing the integral transforms we will now
recall this construction briefly. Let us first define what we mean by
graphs.

\begin{definition}  A graph on $\Sigma$ is a finite subset
$\g\subset\E$ such that $(i)$ two different edges, $e_1, e_2 \ : \ e_1
\neq e_2 $ and $ e_1 \neq e_2^{-1}$, of $\g$ meet, if at all, only at
one or both ends and $(ii)$ if $e \in \g$ then $e^{-1} \in \g$.
\end{definition}

The set of all graphs in $\Sigma$ will be denoted by $\Gra(\S)$.  In
$\Gra(\S)$ there is a natural relation of partial ordering $\ge$, \be
\g'\ \ge\ \g \ee whenever every edge of $\g$ is a path product of
edges associated with $\g'$.  Furthermore, for any two graphs
$\gamma_1$ and $\g_2$, there exists a $\g$ such that $\g \geq \g_1$
and $\g \ge \g_2$, so that $(\Gra(\S), \ge)$ is a directed set.

Given a graph $\gamma$, let ${\cal A}_{\gamma}$ be the associated
space of assignments (${\cal A}_{\g} = \{A_{\g} | A_{\g}: \g
\rightarrow G \}$) of group elements to edges of $\gamma$, satisfying
$A_\g(e^{-1}) = A_\g(e)^{-1}\mbox{ and }A_\g(e_1\circ e_2)=A_\g(e_1)
A_\g(e_2)$, and let $p_{\gamma} : \o{{\cal A}}
\rightarrow {\cal A}_{\gamma}$ be the projection which restricts
$\o{A} \in \o{\cal A}$ to $\g$. Notice that $p_\g$ is a surjective
map.  For every ordered pair of graphs, $\g'\ge\g$, there is a
naturally defined map
\be
\label{0.15} p_{\g\g'}\ :\ \A_{\g'}\
\rightarrow \ \A_\g, \ \ {\rm such}\ \ {\rm that}\ \ p_{\g}\
= \ p_{\g\g'} \circ p_{\g'}.
\ee
With the same graph $\g$, we also associate a group $\G_\g$ defined by
\be
\G_\g\ :=\ \{g_{\g} |g_{\g} : V_{\gamma} \rightarrow G\}
\ee
where $V_{\g}$ is the set of {\it vertices} of $\g$; that is, the set
$V_\g$ of points lying at the ends of edges of $\g$.  There is a
natural projection $\Gb\ \rightarrow\ \G_\g$ which will also be
denoted by $p_\g$ and is again given by restriction (from $\S$ to
$V_\g$). As before, for $\g'\ge\g$, $p_{\g}$ factors into $p_{\g}\ = \
p_{\g\g'} \circ p_{\g'}$ to define
\be
\label{0.17}
p_{\g\g'}\ :\ \G_{\g'}\ \rightarrow \G_\g.
\ee
Note that the group $\G_\g$ acts naturally on $\A_\g$ and that this
action is equivariant with respect to the action of $\Gb$ on $\Ab$ and
the projection $p_\g$. Hence, each of the maps $p_{\g\g'}$ projects to
new maps also denoted by
\be
\label{0.18}
p_{\g\g'} :\ \A_{\g'}/{\cal G}_{\g'}\
\rightarrow \ \A_{\g}/{\cal G}_{\g}.
\ee

We collect the spaces and projections defined above into a (triple)
projective family $(\A_\g, \G_\g, \A_\g/\G_\g,
p_{\g\g'})_{\g,\g'\in\Gra(\S)}$.  It is not hard to see that $\Ab$ and
$\Gb$ as introduced above are just the projective limits of the first
two families.  Finally, the quotient of compact projective limits is
the projective limit of the compact quotients \cite{MM,AL3},
\be
\label{B}
\Ab/\Gb\ =  \agb \ .
\ee
Note however that the projections $p_{\g\g'}$ in (\ref{0.15}),
(\ref{0.17}) and (\ref{0.18}) are different from each other and that
the same symbol $p_{\g\g'}$ is used only for notational simplicity;
the context should suffice to remove the ambiguity.  In particular,
the properties of $p_{\g\g'}$ in (\ref{0.17}) allow us to
introduce a group structure in the projective limit $\Gb$ of $(\G_\g,
p_{\g\g'})_{\g,\g'\in\Gra(\S)}$ while the same is not possible for the
projective limits $\Ab$ and $\agb$ of $(\A_\g,
p_{\g\g'})_{\g,\g'\in\Gra(\S)}$ and $(\A_\g/\G_\g,
p_{\g\g'})_{\g,\g'\in\Gra(\S)}$ respectively.

The $\star$-algebra of {\it cylindrical functions} on $\Ab$ is defined
to be the following subalgebra of continuous functions
\be
 \Cyl(\Ab)\ = \bigcup_{\g\in\Gra(\S)}(p_{\g})^*
C(\A_\g).
\ee
$\Cyl(\Ab)$ is dense in the $C^\star$-algebra of all continuous
functions on $\Ab$. The $\star$-algebra $\Cyl(\agb)$ of cylindrical
functions on $\B$ coincides with the subalgebra of $\Gb$-invariant
elements of $\Cyl(\Ab)$.

Finally, let us turn to the analytic extensions. Since the projections
$p_{\g\g'}$ (in (\ref{0.15}) and (\ref{0.17})) are analytic, the
complexification $G^\Co$ of the gauge group $G$ leads to the
complexified projective family $(\A^\Comp_\g, \G^\Comp_\g,
p^\Comp_{\g\g'})_{\g,\g'\in\Gra(\S)}$.  Note that the projections
$p^{\Co}_\g : \Ab^\Comp \rightarrow \A_\g^\Comp$ maintain
surjectivity. The projective limits $\Ab^\Comp$ and $\Gb^\Comp$ are
characterized as in (\ref{10}) and (\ref{11}) with the group $G$
replaced by $G^{\Comp}$. Since $G^{\Co}$ is non-compact, so will be
the spaces $\Ab^\Comp$ and $\Gb^{\Comp}$.  The algebra of cylindrical
functions is defined as above with $\A^\Co_\g$ substituted for
$\A_\g$.  However these functions may now be unbounded and
$C(\Ab^\Co)$ is not a $C^\star$ algebra.

There is a natural notion of an analytic cylindrical function on $\Ab$
and a holomorphic cylindrical function on $\Ab^\Comp$:

\begin{definition} A cylindrical function $f=f_\g \circ
p_{\g}$ ($f^{\Co} = f^{\Co}_\g \circ p^{\Co}_{\g}$) defined on $\Ab$
($\Ab^\Comp$) is real analytic (holomorphic) if $f_\g$ ($f^{\Co}_\g$)
is real analytic (holomorphic).
\end{definition}

In the complexified case the formula $\ov {\A^\Co/\G^\Co} =
\Ab^\Co/\Gb^\Co$ has not (to the authors' knowledge) been
verified, but the natural isomorphism between $\Cyl(\ov
{\A^\Co/\G^\Co})$ and the algebra of all the $\Gb^\Co$ invariant
elements of $\Cyl(\Ab^\Co)$ continues to exist.  We shall extend it to
define cylindrical holomorphic (analytic) functions on $\ov
{\A^\Co/\G^\Co}$ ($\ov{\A/\G}$) to be all the $\Gb^\Co$ ($\Gb$)
-invariant cylindrical holomorphic (analytic) functions on $\Ab^\Co$
($\Ab$).

\end{subsection}

\begin{subsection} {Measures on $\ov \A$}
\label{meas}

We will now apply to $\Ab$ the standard method of constructing
measures on projective limit spaces using consistent families of
measures (see e.g. \cite{Ya}).

Let us consider the projective family
\be
(\A_\g, \p_{\g \g'})_{\g, \g'\in {\rm{Gra}}(\Sigma)}
\label{321}
\ee
discussed in the last section and let
\be
(\A_\g, \mu_\g,
\p_{\g \g'})_{\g, \g' \in {\rm{Gra}}(\Sigma)} \label{322}
\ee
be a projective family of measure spaces associated with
(\ref{321}); i.e., such that the measures $\mu_\g$ are
(signed) Borel measures on ${\cal A}_\g$ and satisfy the
consistency conditions
\be
\label{consistent} (p_{\g\g'})_*
\mu_{\g'} = \mu_\g \quad for \ \g' \geq \g \ .
\ee
Every projective family of measure spaces defines a cylindrical
measure. To see this, recall first that a set $C_B$ in $\ov {\cal A}$
is called a cylinder set with base $B \subset {\cal A}_\g$ if
\be
C_B = p_\g^{-1}(B) \ , \label{324}
\ee
where $B$ is a Borel set in $\A_\g$. Hence, given a projective family
$\mu_\gamma$ of measures, we can define a cylindrical measure $\mu$ on
$(\overline {\cal A}, {\cal C}_{\overline {\cal A}})$, through
\be
\mu \ : \ p_{\g*} \mu = \mu_\g \ ,
\label{323}
\ee
where ${\cal C}_{\ov {\cal A}}$ denotes the algebra of cylinder sets
on $\ov {\cal A}$.  For a consistent family of measures
$\mu=(\mu_\g)_{\g \in {\rm Gra}(\Sigma)}$ to define a cylindrical
measure $\mu$ that is extendible to a regular ($\sigma$-additive)
Borel measure on the Borel $\sigma$-algebra ${\cal B} \supset {\cal
C}_{\o\A}$ of $\ov \A$ it is necessary and sufficient that the
functional
\be
\ f \mapsto \int d\mu f \  , \ \ f \in \Cyl(\Ab)
\ee
be bounded. This integral is bounded if and only if the family of
measures $ (\mu_\g)_{\g \in {\rm Gra}(\Sigma)}$ is uniformly bounded
\cite{Ba1}; i.e., if and only if $\mu_\g$ considered as linear
functionals on $C(\A_\g)$ satisfy \be || \mu_\g || \leq M \
\label{325} \ee for some $M >0$ independent of $\g$. (If all the
measures $\mu_\g$ are positive then (\ref{325}) automatically holds
\cite{AL1,Ba1}).

{}From now on, all measures $\mu$ on $\ov \A$ will be assumed to
be regular Borel measures unless otherwise stated. It follows from
section 3.1 that every such measure $\mu$ on $\ov \A$ induces a
(regular Borel) measure $\mu'$  on $\agb$
\be
\mu' = \pi_* \mu      \ , \label{326}
\ee
where $\pi$ denotes the canonical projection, $\pi \ : \ \ov \A \
\rightarrow \ \agb$.

The $C^{\omega}$-diffeomorphisms $\varphi$ of $\Sigma$ have a natural
action on $\ov \A$ induced by their action on graphs.  This defines an
action on $C(\ov \A)$ and on the space of measures on $\o\A$ (equal to
the topological dual $C'(\ov \A)$ of $C(\ov \A)$). Diffeomorphism
invariant measures on $\agb$ were studied in \cite{AI}-\cite{Ba1}.  We
will denote the group of $C^\omega$-diffeomorphisms of $\Sigma$
by ${\rm Diff}(\Sigma) $.

A natural solution of conditions (\ref{consistent}) is the one
obtained by taking $\mu_\g$ to be the pushforward of the normalized
Haar measure $\mu_H^{E_\g}$ on $G^{E_\g}$ with respect to
$\psi_\g^{-1}$ where $\psi_{\gamma}: {\cal A}_{\gamma} \rightarrow
G^{E_{\gamma}}$ is a diffeomorphism
\be \label{psi} \psi_{\gamma} : A_{\gamma} \mapsto
(A_{\gamma}(e_1),...,A_{\gamma}( e_{E_{\gamma}} ))
\ee
and
$\{e_1,...,e_{E_\g}\}$ are  edges of $\g$,
such that if (and only if)  $e \in \{e_j\}_{j=1}^{E_\g}$
then  $e^{-1}
\not\in
\{e_j\}_{j=1}^{E_\g}$
\cite{AL1}.
By choosing a different set $\{ \wt e_j \}_{j=1}^{E_\g}$ ($\wt e_j =
e_j^\epsilon , \ \epsilon= 1, -1$) we obtain a different
diffeomorphism $\psi'_\g$. Notice, however, that $\mu_\g$ is well
defined since the map $g \mapsto g^{-1}$ preserves the Haar measure
$\mu_H$ of $G$.  We will refer to the choice of this
$\psi_\g$ as a choice of
orientation for the graph $\g$.  The family of measures $( \mu_\g )_{\g
\in \Gra}$ leads to the measure on $\agb$ denoted in the literature by
$\mu_0$ and for which all edges are treated equivalently.
We will use this measure in section 5.

A method for finding new diffeomorphism invariant measures on $\ov \A$
-- and therefore also on $\agb$ -- was proposed by Baez in \cite{Ba1}.
Since these measures will play an important role in our analysis, we
 now recall some aspects of this method.

\begin{definition} (Baez, \cite{Ba1}). A family $(\mu_\g)_{\g
\in {\rm Gra}(\Sigma)}$ of measures on $\A_\g$ is called
(diffeomorphism) covariant if, for every $\varphi \in {\rm
Diff}(\Sigma)$ and $\g, \g'$ such that $\varphi(\g) \leq
\g'$, we have
\be
(p_{\varphi(\g)\g'})_* \ \mu_{\g'} =  \varphi_* \mu_\g \ .  \label{327}
\ee
\end{definition} As shown in
\cite{Ba1} (Theorem 2), diffeomorphism invariant measures
$\mu$ on $\ov \A$ are in 1-to-1 correspondence with uniformly
bounded covariant families $(\mu_\g)_{\g\in {\rm Gra}(\Sigma)}\ $.
Note that a covariant family is automatically consistent; i.e., it
satisfies (\ref{consistent}).

Baez's strategy is to solve the covariance conditions by appropriately
choosing measures $m_v$ associated with different vertex types $v$.
(Each vertex type is an equivalence class of vertices where two are
equivalent if they are related by an analytic diffeomorphism of
$\Sigma$.)  The number $n_v$ of edge ends incident at $v$ is called
the valence of the vertex. Thus, any edge with both ends at $v$ is
counted twice. For each vertex $v$, the measure $m_v$ is a measure
for $n_v$ $G$-valued random variables $(g_{v1},...,g_{vn_v})$, one for
each of the $n_{v}$ edge ends at $v$.  When applied to the entire
graph, this procedure assigns two random variables $(g_{ea},g_{eb})$
to each of the $E_{\g}$ edges $e \in \g$,
where the variable $g_{ea}$ ($g_{eb}$) corresponds
to the vertex at the beginning (end) of the edge.
We will find it convenient
to alternately label the random variables by their association with
vertices and their association with oriented edges and to denote the map
induced by this
relabelling as $r_\g\; :\; G^{2E_\g}\to G^{2E_\g}$. Given $m_v$ for every
vertex type $v$, we define $\mu_\gamma$ as follows (for a more
detailed explanation see \cite{Ba1}):
\begin{equation}
\label{meas def}
\int_{\A_\g}\;f_\g(A_\g)\;
d\mu_\g(A_\g):=\int_{G^{2E_\g}}(f_\g\circ\psi_\g^{-1}
\circ\phi_{\gamma})\prod_{v\in V_{\g}}
dm_{v}(g_{v1},...,g_{vn_{v}}) \end{equation} where $\psi_\g$
is as in (\ref{psi}) and $\phi_{\gamma}: G^{E_{\gamma}}
\times G^{E_{\gamma}} \rightarrow G^{E_{\gamma}}$ is the map
\be
\label{phi} \phi_{\gamma}:
[(g_{1a},...,g_{E_{\gamma}a}),(g_{1b},...,g_{E_{\gamma}b})]
\mapsto
(g_{1a}g_{1b}^{-1},...,g_{E_{\gamma}a}g^{-1}_{E_{\gamma}b}).
\ee
We will refer to the associated family of measures
$\prod_{v \in V_\g} dm_v(g_{v1},...,g_{vn_v})$ on $G^{2E_\g}$
as $d\mu'_\g$. Notice that (\ref{meas def}) is well defined
because the map (with labelling given by the association
of the random variables with the vertices (!))
\be
\psi_\g^{-1} \circ \phi_\g \circ r_\g \ : \ G^{2E_\g} \ \rightarrow \
{\cal A}_\g
\ee
does not depend on the orientation chosen on the graph,
even though $\psi_\g$, $\phi_\g $ and $r_\g$ do.

The measure $m_v$ has then to satisfy:

(i) If some diffeomorphism induces an inclusion $i$ of $v$
into the vertex $w$, then there is an associated projection
$\pi_i:G^{n_w} \rightarrow G^{n_v}$ acting on the
corresponding random variables.  The measure $m_v$ should
coincide with the pushforward of $m_w$: \begin{equation}
\pi_i^* m_w=m_v \label{328} \end{equation}

(ii) In order to consider embeddings of graphs \[ \g' \geq
\varphi(\g) \] for which several edges of $\g'$ may join to
form in a single edge of $\varphi(\g)$, Baez defines an {\it
arc} to be a valence $2$ vertex for which the two incident
edges join at the arc to form an analytic edge.  He then
proposes the condition that for each valence-1 vertex $v$
connected to an arc $a$ by an edge $e$ (for which the
associated random variables $(g_{ve},g_{ae},g_{ae'})$ have
the distribution $m_v\otimes m_a$) , we have
\begin{equation}
p_{a*}(m_v\otimes m_a)=m_v \label{329} \end{equation} where
$p_a(g_{ve},g_{ae},g_{ae'})=g_{ve}^{-1} g_{ae} g_{ae'}^{-1}$.

In \cite{Ba1} new solutions to conditions (\ref{328}) and
(\ref{329}) were found that distinguish edges as follows.
Let $m$ be an arbitrary but fixed probability measure on $G$.
If a pair of edges $e$ and $f$ meet at an arc $a$ included in
the vertex $v$, set the corresponding random variables equal:
\be g_{a1} = g_{a2} \ . \ee Otherwise the random variables
$g_{vi}$ are distributed according to the measure $m$. Thus,
\be \label{vert} m_v = \prod_{i=1}^{n_v} dm(g_{vi})
\prod_{j=1}^{A_v}\d(g_{vj}, g_{v(n_v-j+1)}), \label{3210}
\ee where $A_v$ denotes the number of arcs included in $v$
and the edge ends have been labeled so that the arcs are
associated with the random variable pairs
$(g_{vi},g_{v(n_v-i+1)})$. The $\delta$-functions in
(\ref{vert}) correspond to the measure $m$. This procedure
defines a measure $\mu^{(m)}$ on $\o\A$ for each
probability measure $m$ on $G$ and we will refer to such
$\mu^{(m)}$ as the {\it Baez measures} on $\Ab$. These
measures distinguish various $n$-valent vertices $v$ by the
number of arcs they include. Additional
diffeomorphism-invariant measures would be expected to
distinguish vertices by using other diffeomorphism invariant
characteristics.

Because $\Ab^{\Co}$ is not compact, it is more difficult to
define $\sigma$-additive measures on this space than on
$\Ab$.  Thus, we content ourselves with cylindrical measures
$\mu$ on $(\Ab^{\Co},{\cal C}_{\A^{\Co}})$.
Cylindrical measures $\mu^\Co$ on $\Ab^\Co$
are in one-to-one correspondence with consistent families
of measures $(\mu_\g)^\Co_{\g \in {\rm
Gra}(\Sigma)}$ exactly as in (\ref{323})
\be
 \ p^\Co_{\g*} \mu^\Co = \mu^\Co_\g \ .
\label{323b}
\ee
The consistency conditions (\ref{consistent}) and diffeomorphism
covariance conditions (\ref{327})
\be
  (p_{\varphi(\g)\g'})_* \ \mu_{\g'} =
 \varphi_* \mu_\g \ .
\ee
 also preserve their forms
\be
 \label{consistentb} (p^\Co_{\g\g'})_*
\mu^\Co_{\g'} = \mu^\Co_\g \quad for \ \g' \geq \g \
\ee
and
\be
  (p^\Co_{\varphi(\g)\g'})_* \ \mu^\Co_{\g'} =
 \varphi_* \mu^\Co_\g \ for \ \g' \geq \varphi(\g)
\ee
respectively. Therefore, diffeomorphism invariant Baez measures
$\mu^{(m)}$ can  be constructed in the same way starting with
an arbitrary probability measure $m^\Co$ on $G^\Co$. We will use these
measures in section 6.

\end{subsection}
\end{section}

\begin{section} {Coherent state transforms for theories
of connections}

The rest of the paper is devoted to the task of constructing coherent
state transforms for functions defined on the projective limit $\Ab$.
The discussion contained in the last two sections makes our overall
strategy clear: we shall attempt to ``glue'' coherent state
transforms defined on the components ${\cal A}_{\g}$ of $\Ab$ into a
consistent family.  However, since the measure-theoretic results are
not as strong for a non-compact projective family, we must first state
under what conditions a map
\be
\label{z1} Z: \L^2(\overline{\A}, d\mu)\
\rightarrow {\cal C}\{{\cal H}_{\cal C}(\overline{\A^\Comp})
\cap L^2(\A^{\Co}, d\nu)\},
\ee
is to be regarded as a coherent state transform.  Here, ${\cal C}$
indicates completion with respect to the $L^2$ inner product and
${\cal H}_{\cal C}$ is the space of holomorphic cylindrical functions.
The definition of the space $L^2(\o{\A}^{\Co}, \nu)$
also requires some care as $\nu$ is not necessarily
$\sigma$-additive.

We first introduce two definitions:
\begin{definition} \label{gauge cov} A transform (\ref{z1})
is $\o\G$-covariant if it commutes with the action of $\o\G$.
That is, if
\begin{equation} \label{com}
Z((L_{\o{g}})^*(f)) = (L^{\Co}_{\o{g}})^*(Z(f))
\end{equation}
where $(A,\o{g})\mapsto
L_{\o{g}} \ov A  := \ov g \ov A $ stands
for the action of $\Gb$ on $\Ab$
with the  superscript $\Co$ denoting the corresponding
action on $\Ab^\Co$:
\be
(L_{\o{g}}^\Co \o{A}^\Co ) (e_{p_1p_2}) = \o{g}_{p_1}
\o{A}^\Co(e_{p_1p_2})\o{g}_{p_2}^{-1}   \label{comm}  \ .
\ee
and where $*$, as usual, denotes the pullback.
\end{definition}

Note that  in  (\ref{com}) and (\ref{comm}), we have
used the inclusion of $\o\G$ in $\o\G^{\Co}$.

\begin{definition} \label{Zconsist} A family $(Z_\g)_{\g \in
{\rm Gra}(\Sigma)}$ of transforms $Z_{\g} : L^2(\A_\g,
d\mu_\g) \rightarrow {\cal H}(\A^\Co_\g)$ is {\it consistent}
if for every pair of ordered graphs, $\g'\ge\g$, \be
\label{35} Z_{\g'}(f_\g \circ p_{\g\g'}) = Z_\g (f_\g) \circ
p^\Co_{\g\g'}   \   .
\ee \end{definition}
\noindent Notice that the consistency condition is equivalent to
requiring that \be p^*_{\g}f_{\g}\ =\ p^*_{\g'}f_{\g'}\
\Rightarrow\ p^{\Co^*}_{\g}Z_\g(f_\g)\ =
p^{\Co^*}_{\g'}Z_\g(f_{\g'})   \   . \ee

Definitions \ref{gauge cov} and \ref{Zconsist} allow us to
use:
\begin{definition} \label{Coh} For a
measure\footnote{Here we identify  measures on $\Ab$ and
$\o{\A}^\Co$ with the
corresponding consistent  families of measures.}
$\mu=(\mu_\g)_{\g\in\Gra(\S)}$ on $\Ab$ and a
cylindrical measure $\nu =
(\nu_\g)_{\g\in\Gra(\S)}$ on $\Ab^\Comp$, a map
(\ref{z1}) is a coherent transform on $\o{\A}$
 if there is a consistent family
$(Z_\g)_{\g\in\Gra(\S)}$ of coherent transforms (see section 2)
\be
Z_{\g}\ :\ L^2(\A_\g, d\mu_{\g})\
\rightarrow \ {\cal H}(\A^\Comp_\g) \cap L^2({\A}^{\Co}_{\g},
d\nu_\g)
\ee such that, for every cylindrical function
of the form $f=f_{\gamma} \circ p_{\g}$ with $f_{\gamma} \in
L^2(\A_\g,d\mu_{\g})$,
\be
\label{e48}
Z(f)\ =Z_{\g}(f_\g) \circ p_\g^\Co    \   .
\ee
\end{definition}

When $Z$ is an isometric coherent transform, it associates with every
representation $\pi$ of the holonomy algebra on $L_2(\agb,\mu)$ a
representation $\pi^{\Co}$ on $L_2(\agbc,\nu)$ by
\be
\pi^{\Co}(\a^{\Co})=Z \pi(\a) Z^{-1}
\ee
where $\a$ is an arbitrary element of the holonomy algebra. Such $\pi^{\Co}$
are the desired ``holomorphic representations".

Several important remarks concerning the properties of the analytic
extensions are now in order. Suppose that we are given a family of
transforms $(Z_\g)_{\g\in\Gra(\S)}$ as in Definition \ref{Zconsist},
but that equation (\ref{35}) is only known to be satisfied when the
functions are restricted to $\A_{\g}\subset\A^\Co_{\g}$ (for every
possible $\g$). Then, because both functions in (\ref{35}) are
holomorphic on $\A^\Co_{\g}$, (\ref{35}) holds on the entire
$\A^\Co_{\g}$.

In other words, in order to construct a family of transforms
$$Z_{\g}: L^2(\A_\g, d\mu_\g) \rightarrow {\cal H}(\A^\Co_\g) \ ,
$$
which is consistent in the sense of Definition \ref{Zconsist}, it is
sufficient to find a family of maps $R_{\g} : L^2(\A_\g, d\mu_\g)
\rightarrow {\cal H}(\A_\g)$ which satisfies (\ref{35}) (${\cal
H}(\A_\g)$ denotes the space of real analytic functions on $\A_\g$).
The analyticity of each function $R_\g(f_\g)$ guarantees the
consistent holomorphic extension.

Let $R: L^2(\Ab, d\mu) \rightarrow L^2(\Ab, d\mu)$ be
the transform defined by restricting $Z(f)$ to $\Ab \subset
\Ab^{\Co}$.  Note that $\Gb$ acts analytically on the components of
the projective family.  Thus, the image of the subspace of
$\o\G$-invariant functions, with respect to a coherent state transform
on $\Ab$, consists of $\Gb^\Co$-invariant functions on $\Ab^\Co$.
\end{section}

\begin{section}  {Gauge covariant coherent state transforms}
\label{gccst}

We now construct a family $Z_t^l$ (parametrized by $t\in \Rl$ and
a function $l$ of edges) of gauge covariant isometric coherent state
transforms when
the measure $\mu$ on $\Ab$ is taken to be the natural measure
$\mu_0$ (see section \ref{meas}).  The corresponding
$Z^l_{t,\g}$ will be coherent state transforms given by appropriately chosen
heat kernels on $\A_\g\cong G^{E_{\gamma}}$. The measures
$\nu_\g$ on the right hand side of (49) are averaged heat kernel
measures on $(G^\Comp)^{E_{\g}}$ (see Section 2).

The idea is to use a Laplace operator $\Delta^{l}$ on $\Ab$
\cite{AL2}. Our transform will then be defined through convolution
with the fundamental solution of the corresponding heat equation.

The ingredients used to define the Laplacian are the following:

\begin{itemize}

\item[(i)] a bi-invariant metric on $G$ which defines the
Laplace-Beltrami operator $\Delta$;

\item[(ii)] a function $l$ defined on the space $\E$ (see
subsection \ref{031}) of (analytic) edges in $\S$, such that:
\be
l(e^{-1})\ =\   l(e),\ \ l(e)\ \ge\ 0 \ ,  \qquad
l(e_2\circ e_1)\ =\ l(e_2)\ +\ l(e_1) \ ,
  \label{lght}
\ee
whenever $e_2\circ e_1$ exists and
belongs to $\E$ and the intersection of $e_1$ with $e_2$
is a single point.

\end{itemize}

Elementary examples of functions $l$ satisfying (\ref{lght}) are given
by: (a) the intersection number of $e$ with some fixed collection of
points and/or surfaces in $\S$; (b) the length with respect to a given
metric on $\S$.
\medskip

To each graph $\g$ we assign an operator acting on functions
on $\A_\g$ as follows,
\be \label{Laplacian}
\Delta^l_\g\ :=\ l(e_1)\Delta_{e_1}\ +\ ...\ +\ l(e_{E_{\gamma}})
\Delta_{e_{E_{\gamma}}},
\ee
where $e_i, i=1,...,E_{\gamma}$ are the edges of $\g$ and
$\Delta_{e_i}$ denotes the pull back, with respect to $\psi^*_\g$ (see
(\ref{psi})), of the operator which is the tensor product of $\Delta$,
acting on the $i$th copy of $G$, with identity operators acting on the
remaining copies.
Because $\Delta$ is a quadratic Casimir operator,
$\Delta^l_\g $ is independent of the choice
of orientation for $\g$.
The condition (\ref{lght}) implies that the family
of operators $(\Delta^l_\g)_{\g\in\Gra(\S)}$ is consistent with the
projective family \cite{AL2}
and therefore defines an operator $\Delta^l$
acting on cylindrical functions.  In other words if $f$ is a
cylindrical function represented by a twice differentiable function
$f_\g$ on $\A_\g$, $f_\g\in C^2(\A_\g)$, then
\be
\Delta^l f\ :=\ (\Delta^l_\g f_\g) \circ p_\g
\ee
and the right hand side does not depend on the choice of the
representative $f_\g$ of $f$. (This would not have been the case if we
had followed a more obvious strategy and attempted to define the
Laplacian without the factors $l(e_i)$ in (\ref{Laplacian}).)

\begin{subsection}{Transform and the main result}

Given a function $l\mbox{ on }\cal E$, the gauge covariant coherent state
transform will be defined with the
help of the fundamental solutions to the heat equation on $\Ab$,
associated with $ \Delta^l$:
\be
\label{heG} {\partial \over \partial t} F_t\ =\   {1 \over 2}
\Delta^l F_t   \ .
\ee
The fundamental solution of (\ref{heG}) is given by the family
$(\rho^l_{t, \g})_{\g\in\Gra(\S)}$ of heat kernels for the operators
$\Delta^l_\g$ on $\A_\g(\cong G^{E_{\gamma}})$,
\be
\label{rho} \rho^l_{t, \g}(A_\g)\ = \
\rho_{s_1}(A_\g(e_1))...\rho_{s_{E_{\g}}}
(A_\g(e_{E_{\g}}))   \ ,
\ee
where $s_i=tl(e_i)$ and each of the functions $\rho_{s}(g)$ being the heat
kernel of the
Laplace-Beltrami operator on $G$. In fact the solution of (\ref{heG})
with cylindrical initial condition
$$ F_{t=0} = f^{(0)}_\g \circ p_\g $$
is given by
\be
F_t = \rho^l_{t, \g}\star f^{(0)}_\g  \ ,
\ee
where the convolution is
\ba
(\rho^l_{t, \g}\star f_\g)(A_\g)\ :=\ &&
\int_{G^{E_\g}}\rho^l_{t, \g} (A^h_\g) \nonumber \\ && \times
(f_\g \circ \psi_\g^{-1}) (h_1,...,h_{E_\g})d\mu_H(h_1)...
d\mu_H(h_{E_\g}) \ , \label{e55}
\ea
and $A^h_\g : e_i \mapsto h_i^{-1}A_\g(e_i)$.  Notice that (\ref{rho})
is well defined since the r.h.s. is invariant with respect to the
change $e_i \mapsto e_i^{-1}$. It is also easy to verify, using the
identity
\be
\label{c1} \int_G \rho_t(g'^{-1}g)f(g'^{-1})d\mu_H(g')\ =\ \int_G
\rho_t(g'^{-1}g^{-1})f(g')d\mu_H(g') \ ,
\ee
that the r.h.s. of (\ref{e55}) does not depend on the orientation
chosen for $\g$ (see discussion after (\ref{psi}). Equality (\ref{c1})
follows from the following properties of the heat kernel \cite{Ha}
\be
\rho_t(g^{-1}) = \rho_t(g) \mbox{ and }
\rho_t(g_1g_2) = \rho_t(g_2g_1) \ .
\ee

Let us consider the family of transforms
$R^l_{t, \g}$ :
\be
R^l_{t, \g}(f_\g) = \rho_{t, \g}^l \star f_\g  \ . \label{eft0}
 \ee

Our main result in the present Section will be:

\begin{theorem} \label{th1} The map
\be
Z^l_t \ : \ L^2(\Ab, \mu) \ \rightarrow \ {\cal C}
\{ {\cal H_C}(\Ab^\Co) \cap L^2(\Ab^\Co , \nu_t^{l}) \} \ ,
\label{e*}
\ee
defined on cylindrical functions $f = f_\g \circ p_\g$ as the analytic
continuation of $R^{l}_{t, \g}(f_\g)$ and extended to the whole
of $L^2(\Ab, \mu)$ by continuity is a gauge covariant isometric
coherent state transform.
\end{theorem}

The measure $\nu^{l}_t$ in (\ref{e*}) is defined below in
subsection \ref{s54}.  We will establish Theorem \ref{th1} with the
help of several Lemmas proved in the following three subsections.
\end{subsection}

\begin{subsection} {Consistency}    \label{0cons0}

Let us first show that the family of transforms (\ref{eft0}) defines a
map of cylindrical functions  on $\Ab$.

\begin{lemma} \label{l1}
The family $(R^l_{t, \g})_{\g \in \Gras}$ in (\ref{eft0}) is
consistent.
\end{lemma}

The proof follows from:
\be
f_\g \circ p_\g = f_{\g'} \circ p_{\g'} \ \Rightarrow \
(\rho_{\g , t}^\l \star f_\g ) \circ p_\g =
(\rho_{\g' , t}^\l \star f_{\g'} ) \circ p_{\g'}   \ .
\label{eal20}
\ee
For convenience of the reader we recall from \cite{AL2} the proof of
(\ref{eal20}).  Since for every pair of graphs $\g_1,\g_2$ there
exists a graph $\g_3\ge\g_1,\g_2$, it is enough to prove (\ref{eal20})
for
\be
\g_2\ \ge\ \g_1   \    .
\ee
The graph $\g_2$ can be formed from $\g_1$ by adding additional edges,
and subdividing edges -- each of these steps being applied some finite
number of times.

Thus, we need only to verify the consistency conditions for each of
the following two cases: the graph $\g_2$ differs from $\g_1$ by $(i)$
adding an extra edge to $\g_1$, and $(ii)$ cutting an edge of $\g_1$
in two.

It follows from the construction of the projective family $(\A_\g,
p_{\g\g'})_{\g , \g' \in \Gras }$ and from formula (\ref{rho}), that
(\ref{eal20}) is equivalent to the following equality
\ba \label{c2}
&& \int_{G^2} \rho_r({g'}^{-1}g)\rho_s({h'}^{-1}h)f(g'h')d\mu_H(g')
d\mu_H(h')\
=\  \nonumber   \\
\ &=& \ \int_{G} \rho_{r+s}({g'}^{-1}gh)f(g')d\mu_H(g') \ .
\ea
for any $r, s\ge 0$.  Eq. (\ref{c2}) follows from (\ref{c1}), from the
fact that $L_g^*$ and $R_g^*$ commute with $\rho_t \star$ for all $g
\in G$ and from the composition rule
\be
\rho_r\star\rho_s\star f = \rho_{r+s}\star f   \   .
\ee
We have:
\begin{eqnarray} &{}&\int_{G^2}
\rho_r({g'}^{-1}g)\rho_s({h'}^{-1}h)f(g'h')d\mu_H(g')
d\mu_H(h') \nonumber\\
&=&\ \int_G \rho_r ({g'}^{-1}g)(\rho_s \star L^*_{g'} f)(h)
d\mu_H(g') \nonumber\\ &=&\ \int_G \rho_r ({g'}^{-1}g)(\rho_s \star
R^*_{h} f)(g') d\mu_H(g')
\nonumber\\ &=& (R^*_{h} \rho_r \star \rho_s
\star f)(g)\nonumber\\ &=& (\rho_r \star \rho_s \star f)(gh)\
= \ (\rho_{r+s}\star f)(gh)\nonumber\\ &=&\int_{G}
\rho^\Comp_{r+s}({g'}^{-1}gh)f(g')d\mu_H(g')  \ .
\end{eqnarray}
This completes the proof of (\ref{eal20}) and therefore also of Lemma
\ref{l1}.
\bigskip

According to Lemma \ref{l1}, given a cylindrical function $f=f_\g
\circ p_{\g}$ we have a well defined ``heat evolution",
\be R_t^l  (f) \ :=\ R^l_{t, \g} (f_\g)  \circ p_\g  \ .
\ee
Notice that from Section 2 it follows that for any $f_\g \in
L^2(\A_\g, d\mu_{0,\g})$ the convolution $\rho^l_{t, \g}\star f_\g =
f_\g \star \rho_{\g , t}^l $ is a real analytic function.


We define a coherent state transform on each $\A_\g$ through
\be
\label{Z_t}
(Z^l_{t, \g}f_\g)(A_\g^{\Co})\ :=\ (\rho^{l\Comp}_{t, \g}\star
f_\g)(A_\g^{\Co})   \    ,
\ee
where $\rho_{t, \g}^{l\Co}$ is the analytic continuation of
$\rho^l_{t, \g}$ from $\A_\g$ to $\A^{\Co}_\g$
\cite{Ha}.
According to Lemma \ref{l1} and the remarks after Definition
\ref{Coh}, the family of transforms  \break
$(Z^l_{t, \g})_{\g\in\Gra(\S)}$ is consistent in the sense of
Definition \ref{Zconsist}. Hence, we may define the transform for each
square-integrable cylindrical function $f=f_\g \circ p_{\g}\in
\Cyl(\Ab)$: \be \label{Z} Z^l_t(f)\ :=Z^l_{t, \g}[f_\g] \circ
p^{\Co}_\g \
\ee
which maps the space of $\mu_0$-square-integrable cylindrical
functions on $\Ab$ into the space of cylindrical holomorphic functions
on $\Ab^\Co$.

\end{subsection}


\begin{subsection} { Measures on $\o  {\cal A}^\Co$} \label{s54}
\bs
\medskip

Consider the averaged heat kernel measure $\nu_t$ (\ref{avhk})
defined on the complexified group $G^\Comp$ and the associated family
of measures $(\nu^l_{t,\g})_{\g\in\Gra(\Sigma)}$ on the spaces
$\A^\Co_\g$ :
\be
\label{nu}
d\nu^l_{t,\g}(A^\Co_\g)\ :=\
d\nu_{l(e_1)t}(A^\Co_\g(e_1)) \otimes ... \otimes
d\nu_{l(e_{E_\g})t}(A^\Co_\g(e_{E_\g})) \ .
\ee
It follows automatically from \cite{Ha} that the transform $Z^l_{t,
\g} \ : \ L^2(\A_\g,d\mu_{\g,AL}) \rightarrow {\cal H} (\A_\g) \cap
L^2(\A^\Co_\g, d\nu_{t,\g}^l)$ is isometric.  Isometry of the
transforms $Z^l_{t, \g}$ implies the following equality for all
square-integrable holomorphic functions $f_{1_\g}, \ f_{2_\g}$
and all $\g' \geq \g$
\be
\int_{\A^\Co_\g} \ov{f_{1_\g}(A_\g^\Co)} f_{2_\g}(A_\g^\Co)
d\nu^l_{t,\g} =
\int_{\A^{\Co}_{\g'}} \ov{(f_{1_\g} \circ p_{\g\g'}^\Co) (A_{\g'}^{\Co})}
(f_{2_\g} \circ p_{\g\g'}^\Co) (A_{\g'}^\Co)
d\nu^l_{t,\g'} \ .                         \label{0071}
\ee
{}From the arbitrariness of $f_{1_\g}$ and $f_{2_\g}$ we will
conclude that the family
\break
$\{\nu_{t,\g}^{l,\Co} \}_{\g,\g' \in \Gra(\S)}$ is consistent and
therefore defines a cylindrical measure on $\Ab^{\Co}$ which will be
denoted by $\nu_t^{l}$.

To see this let $i \ : \ \wh G^\Co \ \rightarrow \ \Co^N$ be an
analytic
immersion of $\wh G^\Co := G^\Co \times ... \times G^\Co$ into $\Co^N$
for sufficiently large $N$. A Borel probability measure $\mu^\Co$ on
$G^\Co$ defines a Borel probability measure $i_* \mu^\Co$ on $\Co^N$
(supported on $i(\wh G^\Co)$) through
\be
\int_{\Co^N} f d(i_* \mu^\Co) : = \int_{\wh G^\Co} i^* (f)   d\mu^\Co  \  .
\ee
Consider the analytic functions $i^* (F_\l)$ on $\wh G^\Co$, where
\be
F_\l (z) = e^{\l z} \ , \quad \l , z \in \Co^N \ , \quad \l z :=
\sum_{j=1}^N \l_j z_j   \ .
\ee
For every $\d_1, \d_2 \in \Rl^N$ we
choose $\l_1 = - 1/2(\d_2 + i \d_1)$
and $\l_2 = - \l_1$ so that
\be
(\overline F_{\l_1} F_{\l_2})(x,y) = e^{i(\d_1 x + \d_2 y)}    \  ,
\ee
where $z = x+iy$.
Then
\be
\chi_{\mu^\Co}(\d_1, \d_2)  : = \int_{\Rl^{2N}} e^{i(\d_1 x + \d_2 y)}
d(i_* \mu^\Co)
= \int_{\wh G^\Co} \overline {i^*( F_{\l_1})}i^*(F_{\l_2}) d \mu^\Co
\ee
is the Fourier transform of the measure $i_*\mu^\Co$ on $\Rl^{2N}$,
which, according to the Bochner theorem, completely determines
$i_*\mu^\Co$ and therefore also $\mu^\Co$. Thus, (\ref{0071})
implies that $(p_{\g \g'}^\Co)_* \nu_{t, \g'}^{l}$ and
$ \nu_{t, \g}^{l} $ in fact agree as Borel measures on
$\A_\g^\Co$.

\end{subsection}

\begin{subsection} {Gauge covariance}

Here we complete the proof of Theorem \ref{th1}.
\bigskip

We only need to establish:

\begin{lemma} \label {Rcov} $R$ commutes with the
action of $\o\G$ on $L^2(\o\A,d\mu_0)$. \end{lemma}

In the proof, $g, g_a, g_b$, and $\psi_\g(A_\g)$ will be elements of
$G^{E_\g}$ and we define multiplication of $E_\g$-tuples
component-wise; i.e., $(g_ag_b)_i = (g_a)_i (g_b)_i$.

{\it Proof of Lemma \ref{Rcov}}. \ For cylindrical $f = f_\g
\circ p_\g$ and $ \o g \in \o\G$, let  $g_a,g_b \in G^{E_\g}$
be given by $(g_a)_i := \o{g}(p_{ia})$ and $(g_b)_i := \o{g}(p_{ib})$,
where $p_{ia}$ and $p_{ib}$ are the initial and final points of the
edge $e_i$ associated with a fixed choice of orientation on $\g$.
Then,
\ba
R^l_t[f](\o{g}[{A}_\g]) &=& (\rho_{t, \g} \star
f_\g)(\o{g}[A_\g]) =
\nonumber \\
&=& \int_{G^{E_\g}} (f_\g
\circ \psi^{-1}_\g )
(g g_a \psi_\g(A_\g)g_b^{-1}) \prod (\rho_t
d\mu_H)(g) =
\\
 &=& \int_{G^{E_\g}}( f_\g
\circ \psi^{-1}_\g)
 (g_a g
\psi_\g(A_\g)g^{-1}_b) \prod (\rho_t d\mu_H)(g) =
\nonumber \\
&=& R^l_t [\o{g}^*(f)](\o{A})    \nonumber
\ea
since the measure is conjugation invariant.  Note that this is a
consequence of the $\o\G$-invariance of $\Delta^l$.

Finally, note that since the transform (\ref{Z}) depends on the path
function $l$, it fails to be diffeomorphism covariant.

\end{subsection}

\end{section}

\begin{section} {Gauge and diffeomorphism covariant coherent
state transforms} \label{DI}

In this Section, we introduce a coherent state transform that is both
gauge and diffeomorphism covariant.  This new transform will be based
on techniques associated with the Baez measures  and we  recall from
subsection \ref{meas} that, given any Baez measure $\mu^{(m)}$
on $\Ab$ and the corresponding measures $\mu^{(m)}_{\gamma}$ on
${\cal A}_{\gamma}$, we may write (\ref{meas def}) as
\be
\int_{{\cal A}_{\gamma}} f_{\gamma} d\mu^{(m)}_\gamma =
\int_{G^{E_{\gamma}} \times G^{E_{\gamma}}} f_{\gamma} \circ
\psi^{-1}_{\gamma} \circ \phi_{\gamma} \ d\mu^{(m)}_\gamma{}'
\   .
\ee
{}From (\ref{vert}), each $d\mu^{(m)}_\gamma{}'$ is a
product of measures $dm$ on $G$ and delta functions with
respect to these measures.  The arguments of the delta
functions are pairs of coordinates and no coordinate appears
in more than one delta function. Specifically, this is true
for the Baez measure $\tilde{\mu}_0\equiv\mu^{(\mu_H)}$
constructed from the Haar measure $m=\mu_H$ on $G$.

\begin{subsection} {The transform and the main result}
\label{afot}
Let us fix a measure  $\nu$ on $G^{\Co}$ that satisfies
the conditions listed in Section \ref{Hall} for the existence
of the Hall transform $C_{\nu}$. Given $\nu$
we have on $G$ a generalized
heat-kernel measure $d\rho = \rho_{\nu}d\mu_H$
used in the Hall transform (\ref{HT}) from $L^2(G,\mu_H)$ to
$L^2(G^{\Co},\nu) \cap {\cal H}(G^\Co)$.

Our transform will be defined as follows.  Given some
$\o{A}_0 \in \o{\cal A}$ and the corresponding $A_{0,\gamma}
\in {\cal A}_{\gamma}$, let $\phi_{\o{A}_0,\gamma}: G^{E_{\gamma}}
\times G^{E_{\gamma}} \rightarrow G^{E_{\gamma}}$ be the map
\ba
\phi_{\o{A}_0,\gamma}  \  &:&  \
[(g_{1a},...,g_{E_{\gamma}a}),(g_{1b},...,g_{E_{\gamma}b})]
 \nonumber \\
&& \mapsto \ (g_{1a}\o{A}_0(e_1)g_{1b}^{-1},...,
g_{E_{\gamma}a}\o{A}_0(e_{E_{\gamma}}) g_{E_{\gamma}b}^{-1})\ .
\ea
Note that $\phi_{\o{A}_0,\gamma}$ depends on $\o{A}_0$ only through
$A_{0,\gamma}$ and that if $\o{A}_0$ is the trivial connection $\o{1}$
(for which $\o{1}(e) = 1_G$ for any $e \in {\cal E}$) then
$\phi_{\o{1},\g} = \phi_\g$ of (\ref{phi}).

 For $f: \o{\cal A}
\rightarrow \Co$ such that $f = f_{\gamma} \circ p_{\gamma}$,
we would like to define  $R(f): \o{\A} \rightarrow
\Co$ through $R  (f) =
 R_\g (f_{\gamma}) \circ
p_{\gamma}$, where
\be
\label{trans}
 R_\g ( f_{\gamma}) (A_{0,\gamma}) =
 \int_{G^{2E_{\gamma}}} f_{\gamma}
\circ \psi_{\gamma}^{-1} \circ \phi_{\o{A}_0,\gamma} d\rho'_{\gamma}\ .
\ee
In (\ref{trans}) $d\rho'_{\gamma}$ is the
measure on $G^{E_{\gamma}} \times G^{E_{\gamma}}$ associated with the
Baez measure $\rho= \mu^{(\rho)}$.
Thus, $d\rho'_{\gamma}$ is a product of
generalized heat kernel measures $d\rho$ and delta-functions with
respect to this measure.
We will show that the map  $R$ is well defined.
Our main result will be

\begin{theorem} \label{dccst} For each $\nu$, there exists
a unique isometric  map
\be
Z \ : \ L^2(\o{\cal A}, \tilde{\mu}_0) \ \rightarrow \
{\cal C}\{ {\cal H}_{\cal C}(\Ab^\Co) \cap L^2(\o{\cal A}^{\Co},
\mu^{(\nu)}) \}   \   ,
\ee
such that, for every  $f\in\Cyl(\Ab)$ and any holomorphic
($L^2$-)representative  of $Z(f)$ with restriction to $\Ab$
denoted by $\wt f$,
the real-analytic function $\wt f $
coincides  $\tilde{\mu}_0$-everywhere with   $R(f)$. The map
$Z$ is a gauge and  diffeomorphism covariant isometric coherent
state transform.
\end{theorem}

\end{subsection}

\begin{subsection}  {Consistency}

As before, it is convenient to break the proof of our theorem into
several parts.  We begin with

\begin{lemma} \label{ucl1}
The family $(R_\g)_{\g \in {\rm Gra}
(\Sigma)}$,
\be
{R}_\g(f_\g)({A}_{0,\g}) =
 \int_{G^{2E_{\gamma}}} f_{\gamma} \circ
\psi_{\gamma}^{-1} \circ \phi_{\o{A}_0,\gamma} d \rho'_{\gamma} \ ,
\ee
is consistent.
\end{lemma}

{\it Proof} \ Suppose that $f:\o{\cal A} \rightarrow \Co$ is
cylindrical  with $f = f_{\gamma_1} \circ p_{\gamma_1}$ and $f =
f_{\gamma_2} \circ p_{\gamma_2}$.  As in Section \ref{gccst}, it is enough to
consider the case $\g_2 \geq \g_1$.

We must now establish the conditions $(i),\ (ii)$ listed in the proof
of Lemma \ref{l1} in subsection \ref{0cons0}. The first case is
straightforward.
Indeed $f_{\gamma_2} = f_{\gamma_1} \circ p_{\gamma_1 \gamma_2}$
depends only on those edges that actually lie in $\gamma_1$.
Integration over the other variables in the measure
$d\rho'_{\gamma_2}$ simply yields the measure $d\rho'_{\gamma_1}$ as
in the usual Baez construction. Thus,
$R_{\g_2}({f}_{\gamma_2}) =
R_{\g_1}
({f}_{\gamma_1}) \circ p_{\gamma_1 \gamma_2}$.

We now address $(ii)$.  Suppose that $\gamma_2$ is just $\gamma_1$
with the edge $e_0 \in \gamma_1$ split into $e_1$ and $e_2$ at the
vertex $v$.  Let $e_1$, $e_2$ have orientations induced by $e_0$.
Without loss of generality, let $e_1 \circ e_2 = e_0$. Then we have
\be
 R_{\g_2}({f}_{\gamma_2}) (A_{0,{\gamma_2}}) =
\int_{G_a^{E_{\gamma_2}} \times G_b^{E_{\gamma_2}}}
(f_{\gamma_2} \circ \psi^{-1}_{\gamma_2}) (g_{1a}
\o{A}_0(e_1) g_{1b}^{-1},...) d\rho'_{\gamma_2} \    ,
\ee
where the
$g_{ia}$ are coordinates on $G_a^{E_{\gamma_2}}$ and the
$g_{ib}$ are coordinates on $G^{E_{\gamma_2}}_b$.  Since
$f_{\gamma_2} = f_{\gamma_1} \circ p_{\gamma_1 \gamma_2}$,
$(f_{\gamma_2} \circ \psi^{-1}_{\gamma_2})
(g_1,...,g_{E_{\gamma_2}}) = (f_{\gamma_1} \circ
\psi^{-1}_{\gamma_1}) (g_1g_2,g_3,...,g_{E_{\gamma_2}})$, it
follows that
\ba
&&  R_{\g_2}({f}_{\gamma_2})(A_{0,\gamma_2})    =
\nonumber \\ &=& \int_{G_a^{E_{\gamma_2}} \times G_b^{E_{\gamma_2}}}
(f_{\gamma_1} \circ
\psi^{-1}_{\gamma_1})
(g_{1a}\o{A}_0(e_1)g_{1b}^{-1}g_{2a}\o{A}_0(e_2)
g_{2b}^{-1},g_{3a}\o{A}_0(e_3)g_{3b}^{-1},...
\nonumber \\
&& g_{E_{\gamma}a}\o{A}_0(e_{E_{\gamma}})g_{E_{\gamma}b}^{-1})
 \times \delta(g_{1b},g_{2a})
d{\rho}(g_{1b})d{\rho}(g_{2a})
\\ &&
d\rho'_{\gamma_1}[(g_{1a},g_{3a},...,
g_{E_{\gamma}a}),(g_{2b},g_{3b},...,g_{E_{\gamma}b})]  = \nonumber \\
&=& ( R_{\g_1}({f}_{\gamma_1})  \circ \psi_{\gamma_1}^{-1})
(\o{A}(e_1)\o{A}(e_2), \o{A}(e_3),...,\o{A}(e_{E_{\gamma}}))  =
\nonumber \\ &=&
( R_{\g_1}( {f}_{\gamma_1}) \circ p_{\gamma_1 \gamma_2})
(A_{0, \g_2})    \    .
\nonumber
\ea

This is enough to show consistency so that
  the family $(R_\g)_{\g\in\Gra(\S)}$ defines
unambiguously a map
$R\ :\ \Cyl(\Ab)\cap L^2(\Ab, \tilde{\mu}_0) \rightarrow
\Cyl(\Ab)$.

\end{subsection}

\begin{subsection}{Extension and isometry}       \label{0iso}
For a general $f\in \Cyl(\Ab)\cap L^2(\Ab, \tilde{\mu}_0)$, the function
$R(f)$ may not be real-analytic on $\Ab$. However, there still
exists a natural ``analytic extension''
 of $R(f)$ to a unique element of  $L^2({\cal A}^{\Co},
\mu^{(\nu)})$ that can be briefly defined as follows. The function
$R(f)$ is
real-analytic when restricted to a subspace of $\Ab$ carrying the
support of the Baez measure; on the other hand, the
complexification of this subspace contains the support of the Baez measure
in $\Ab^\Co$. This is sufficient for the extension of $R(f)$
to exist and be unique (in the sense  of $L^2$ spaces).

To define the extension more precisely,
let us
 first
express the Baez integral in a  more convenient  form.
Given an oriented  graph $\g$, consider  $\A_\g$, $\A_\g^\Co$ and
the corresponding maps $\psi^{-1}_\g \circ \phi_\g:G^{E_\g}\times G^{E_\g}
\rightarrow \A_\g$ as well as the complexification
 $\psi^{\Co-1}_\g \circ \phi^\Co_\g:G^{\Co E_\g}\times G^{\Co E_\g}
\rightarrow \A^\Co_\g$.  In what follows, all the functions on $\A_\g$
($\A_\g^\Co$)
shall be identified with their pullbacks to  the corresponding
$G^{E_\g}\times G^{E_\g}$ ($ G^{\Co E_\g}\times G^{\Co E_\g} $). Since the
delta-functions in the Baez measure identify some pairs $(g_{ia},g_{jb})$ of
variables,  for some $E_{\gamma} \leq k_{\gamma} \leq 2E_{\gamma}$,
they define embeddings
\ba
\lambda_{\gamma}\ &:&\ G^{k_{\gamma}}\ \rightarrow\ G^{E_{\gamma}} \times
G^{E_{\gamma}}  \nonumber          \\
\lambda^\Co_{\gamma}\ &:& \ G^{\Co k_{\gamma}}\ \rightarrow\ G^{\Co E_{\gamma}}
\times G^{\Co E_{\gamma}}   \ ,
\ea
where $\lambda^\Co_\g$ is the complexification of $\lambda_\g$ and both are
insensitive to the choice of measure on $G$ used to define the Baez measure.
(Note that the maps $\lambda$ and $\psi^{-1} \circ \phi \circ\lambda$ do
not
depend on the choice of an orientation of $\g$.)

Suppose that we wish to compute the integral of some  $f =
f_{\gamma} \circ p_{\gamma}\in\Cyl(\Ab)$  with respect to
$\tilde{\mu}_0 = \mu^{(\mu_H)}$  or $f = f_{\gamma} \circ p^\Co_{\gamma}\in
\Cyl(\Ab^\Co)$  with respect to $\mu^{(\nu)}$.
 Then, we may use these embeddings  to write the integrals as
\be \label{int}
\int_{\o{\cal A}} f \
d\tilde{\mu}_0 =  \int_{G^{k_{\gamma}}}
f_{\gamma}\circ \psi^{-1}_{\gamma} \circ \phi_{\gamma} \circ
\lambda_{\gamma} \prod d\mu_H
\ee
\be
\label{intC}
\int_{\o{\cal A}^\Co} f \ d\mu^{(\nu)} = \int_{G^{\Co k_{\gamma}}}
f_{\gamma}\circ \psi^{\Co^{-1}}_{\gamma} \circ \phi^\Co_{\gamma} \circ
\lambda^\Co_{\gamma} \prod d\nu.
\ee
The above formulas show the following statement.

\begin{lemma}\label{rest} Let $f_1, f_2\in \Cyl(\Ab)$; $f_1 =f_2$
$tilde{\mu}_0$-everywhere  if and
only if for a
graph $\g$ such that $f_i= p_\g^* f_{i \g} \ i = 1, 2$, we have
$(\psi^{-1}_\g \circ \phi_\g \circ\lambda_\g)^* f_{1\g} =
(\psi^{-1}_\g \circ \phi_\g \circ\lambda_\g)^* f_{2\g}$  $\prod
d\mu_H$-everywhere (and analogously for the complexified
case). The natural maps
\ba
(\psi^{-1}_\g \circ \phi_\g \circ\lambda_\g)^*\ &:& \
L^2(\A_\g, \tilde{\mu}_{0\g})
 \rightarrow  L^2(G^{k_\g}, \prod \mu_H^\Co) \ ,
\nonumber \\
(\psi^{\Co-1}_\g \circ \phi^\Co_\g \circ\lambda^\Co_\g)^*\ &:& \
L^2(\A^\Co_\g,
\mu^{(\nu)}_{\g})
 \rightarrow
L^2(G^{\Co k_\g}, \prod \nu) \  ,
\ea
 are isometric.
 \end{lemma}

Further, let $C_{(k_{\gamma})}$ be the coherent state transform defined by
Hall from $L^2(G^{k_{\gamma}}, \prod_{i=1}^{k_{\gamma}} d \mu_H(g_i))$
to $L^2(G^{\Co^{k_{\gamma}}}, \prod_{i=1}^{k_{\gamma}}
d\nu(g_i^\Co))$. It follows from (\ref{int}, \ref{intC}) that
\ba
[
R_\g({f}_{ \gamma})
\circ \psi^{-1}_{\gamma} \circ \phi_\g
\circ \lambda_\g]
(g^*) &=& \int_{G^{k_{\gamma}}}
[f_{\gamma} \circ \psi^{-1}_{\gamma} \circ \phi_{\gamma}
\circ \lambda_{\gamma}] (g^{-1} g^*) \prod d\rho(g) = \nonumber \\
&=& (C_{(k_\g)}[f_{\g} \circ \psi^{-1}_{\gamma} \circ
\phi_{\gamma} \circ \lambda_{\gamma}])(g^*) \    ,
\label{uc14}
\ea
where $g,g^*, gg^* \in G^{k_{\gamma}}$ and $(gg^*)_i = g_i g^*_i$.
Re-expressing the last result less precisely, the restriction of
$R_\g(f)$ to $G^{k_\g}$ embedded in $G^{E_\g}\times G^{E_\g}$
coincides with the usual Hall transform.
The following Lemma then follows from the results of
\cite{Ha}.

\begin{lemma} \label{restanal}
Let $f_\g$ be a measurable function on $\A_\g$ with
respect to the Baez measure $\tilde{\mu}_{0\g}$; the function $R_\g(f_\g)$
restricted to $\psi_\g^{-1} \circ \phi_\g \circ
\lambda_\g (G^{k_\g})$  is real-analytic.
\end{lemma}

The function $R_\g(f_\g)$ can thus be analytically extended
to a holomorphic function  defined on
$\psi_\g^{\Co -1} \circ \phi_\g^\Co \circ \lambda_\g^\Co(G^{k_\g \Co})$
which, according to Lemma \ref{rest},
 uniquely determines an element $Z_\g(f_\g)$ in
$L^2(\A^\Co_\g, \nu_\g)$.
We have defined  a map $Z_\g$
\be\label{Z}
Z_\g\ :\  L^2(A_\g,\tilde{\mu}_{0\g})\ \rightarrow L^2(A^\Co_\g,
\mu_{\g}^{(\nu)}) \  .
\ee
The consistency of the family of maps $(Z_\g)_{\g\in\Gra(\S)}$ easily follows
from the consistency of $(R_\g)_{\g\in\Gra(\S)}$. Another advantage of
 relating, through   (\ref{uc14}),  $Z_\g$ with
the usual Hall transform $C_{(k_\g)}$ is that
we may again consult Hall's results and note that the map
(\ref{Z}) is an
isometry.
Thus, we have verified the following  Lemma.

\begin{lemma} \label{ZZ}

$ \  $

\noindent(i) The family of maps $(Z_\g)_{\g\in\Gra(\S)}$ (\ref{Z}) is
consistent;

\noindent(ii) The map
\be
Z\ :\ L^2(\Ab, \tilde{\mu}_0)\cap \Cyl(\Ab)\ \rightarrow
L^2(\Ab^\Co,\mu^{(\nu)})
\ee
is an isometry, where $Z(p_\g^*f_\g)\ := Z_\g(f_\g)$.

\end{lemma}

Since cylindrical functions are dense in $L^2(\o{\cal
A},\tilde{\mu}_0)$, it follows that our transform $Z$
extends to
\be
Z \ :\ L^2(\Ab , \tilde{\mu}_0) \rightarrow\
{\cal C} \{ L^2(\Ab^\Co,
{\mu}^{(\nu)}) \}
\ee
as an isometry.
\end{subsection}

\begin{subsection} {Analyticity } \label{RAN}

We have seen that the pullback of $Z_\g (f_\g)$ through the map $
(\psi_\g^\Co)^{-1} \circ  \phi_\g^\Co \circ \lambda_\g^\Co$
may be taken to be holomorphic.  However, we will now show that
this is the case for  $Z(f)$ itself.

\begin{lemma} \label{analyticity2}
 If $f\in\Cyl(\Ab)\cap L^2(\Ab, \tilde{\mu}_0)$ then,

\noindent (i) Any cylindrical function
$f = f_{\gamma} \circ p_{\gamma}$ differs only on a set of
$tilde{\mu}_0$  measure zero from some $f^0 =
f^0_\g \circ p_\g$ such that
$R_\g ({f}^0_{\gamma})$ is real analytic.

\noindent (ii) $Z(f)$ may be represented by
a holomorphic function on $\Ab^\Co$.

\end{lemma}

Note that the second part of the Lemma follows automatically from
part
$(i)$.

For this Lemma, we will use the concept of the {\it Baez-equivalence
graph} $\gamma_E$ corresponding to a graph $\gamma$.  This $\g_E$ is
an abstract graph (a collection of ``edges" and ``vertices" not
embedded in any manifold) formed from the edges of $\gamma$.
However, two
edges in $\gamma_E$ meet at a vertex  if and only if the corresponding
edges join to form an analytic path in $\gamma$.  Since each edge of
$\gamma$ can, at a given vertex, meet at most one other edge
analytically, each vertex in $\gamma_E$ connects at most two edges.
Thus, $\gamma_E$ consists of a finite set of line segments and closed
loops
that do not intersect.
Let us orient the edges of $\gamma_E$ so that, at each vertex,
one edge flows in and one edge flows out.  We will assume that the
edges of $\gamma$ are oriented in the corresponding way.

A graph $\gamma$ for which $\gamma_E$ contains no cycles will be
called {\it Baez-simple}.  To derive Lemma \ref{analyticity2}, we will
also need the following Lemma:

\begin{lemma} \label{simple} Any cylindrical function $f:\o{\cal A}
\rightarrow \Co$ is identical to a function $f^0$ that is cylindrical
over a Baez-simple graph $\g_s$, except on sets of $tilde{\mu}_0$ measure
zero. \end{lemma}

To see this, we construct the Baez-simple graph $\gamma_s$ from
$\gamma$ by removing one edge $e_0^i$ from the $i$th cycle in
$\gamma_E$.  Let $\zeta \ : \ {\cal A}_{\gamma}
\rightarrow {\cal A}_\gamma$ be the map such that
\be
[\zeta(A_{\gamma})](e^i_0) = [\prod_{j=1}^{N_i}
A_{\gamma}(e^i_j)]^{-1} \    ,
\ee
where $e^i_j$ are the other edges in the $i$th cycle and are numbered
from $1$ to $N_i$ in a manner consistent with their orientations.  For
any other edge $e$, let $[\zeta(A_{\gamma})](e) = A_{\gamma}(e)$.

{\it Proof of Lemma \ref{simple}} \ If $f = f_{\gamma} \circ
p_{\gamma}$, let $f^0_{\gamma} = f_{\gamma} \circ \zeta$ and $f^0 =
f^0_{\gamma} \circ p_{\gamma}$ so that $f^0$ is in fact cylindrical
over $\gamma_s$ ($f^0 = f^0_{\gamma_s} \circ p_{\gamma_s}$). Note that
$d\mu'_{\gamma}$ is a product of measures associated with the
connected components of $\gamma_E$ and recall that $f^0_{\gamma}$
differs from $f_{\gamma}$ only in its dependence on edges in cycles of
$\gamma_E$.  For simplicity, let us assume for the moment that
$f_{\gamma}$ in fact depends only on edges that lie in one cycle
$\alpha$ in $\gamma_E$ so that $f_{\gamma} = f_{\alpha} \circ
p_{\alpha \gamma}$ for some $f_{\alpha}: {\cal A}_{\alpha} \rightarrow
\Co$.  Furthermore,

\vbox{

\ba   \label{0079}
&& ||f
- f^0||^2_{L^2,\tilde{\mu}_0} = \nonumber \\ && =
\int_{G_a^{E_{\alpha}} \times G_b^{E_{\alpha}}} |[f_{\alpha}
\circ \psi^{-1}_{\alpha}]
(g_{0a}g^{-1}_{0b},g_{1a}g^{-1}_{1b},...,
g_{(E_{\alpha}-1)a}g^{-1} _{(E_{\alpha}-1)b}) \nonumber \\
 && - [f_{\alpha}\circ \psi_{\alpha}^{-1}]
((\prod_{i=1}^{E_{\alpha}-1} (g_{ia}g^{-1}_{ib}))^{-1},
g_{1a}g^{-1}_{1b},...,g_{(E_{\alpha}-1)a}
g^{-1}_{(E_{\alpha}-1)b})|^2
\\ && \prod_{i=1}^{E_{\alpha} -1}\delta(g_{(i-1)b},g_{ia})
\delta(g_{(E_{\alpha}-1)b}, g_{0a})
\prod_{j=0}^{E_{\alpha}-1} d\mu_H(g_{ia}) d\mu_H(g_{ib}) = 0
\    ,    \nonumber
\ea

}

\noindent so that $f$ and $f^0$ differ only on sets of $\tilde{\mu}_0$
measure zero.  The same is true when $f_{\gamma}$ depends on several
cycles $\alpha_i$.\\

We can also use $\g_E$ to introduce a convenient labelling of the
edges in $\g_s$.  Let $e_{(i,j)}$ be the $j$th edge of the $i$th
connected component of $\gamma_E$, where we again assume that the
edges in the $i$th component are numbered consistently with their
orientations. Note that since $\g_s$ is Baez-simple these components
form open chains with well-defined initial edges $(e_{(i,1)})$ and
final edges $(e_{(i,N_i)})$.

{\it Proof of Lemma \ref{analyticity2}} \ Suppose that there are
$N_{\g_s}$ components of $\g_s$.  Then, from (\ref{meas def}),
(\ref{phi}), and (\ref{vert}) we have
\ba
d\rho'_{\g_s}
&=& \prod_{i=1}^{N_{\g_s}} \Bigg[ d
\rho(g_{(i,1)a}) d\rho(g_{(i,1)b}) \nonumber \\
&& \prod_{j=2}^{N_i}
\delta(g_{(i,j-1)b}, g_{(i,j)a}) d \rho(g_{(i,j)a})
d\rho(g_{(i,j)b}) \Bigg]    \ .
\ea
For $k_{\g_s} =
\sum_{i=1}^{N_{\g_s}} (N_i+1)$, let us now introduce the map
$\sigma_{\o{A},\g_s} : G^{k_{\g_s}} \rightarrow
G^{E_{\g_s}}$ through
\begin{equation}
[\sigma_{\o{A},\g_s}(g)]_{i,j}
= g_{(i,j)} \o{A}(e_{(i,j)}) g^{-1}_{(i,j+1)} \end{equation}
for $g \in G^{k_\g}$, where we have set
$g{(i,1)} =
g_{(i,1)a}$ and $g_{(i,j)} = g_{(i,j-1)b}$ for $j \geq 2$.
Thus, we may write
\begin{equation}
\label{uc22}
R_{\g_s}({f}^0_{\g_s})(A_{\g_s})
= \int_{G^{k_{\g_s}}} [f^0_{\g_s}
\circ \psi^{-1}_{\g_s} \circ \sigma_{\o{A},\g_s}]
\prod_{i=1}^{N_{\g_s}} \prod_{j=1}^{N_i+1} \rho(g_{(i,j)})
d\mu_H(g_{(i,j)})    \     .
\end{equation}

Analyticity of (\ref{uc22}) can
now be shown by making the change of integration
variables:
\begin{equation}
g'_{(i,j)} = g_{(i,j)} \prod_{k=j}^{N_i+1}
A(e_{(i,k)})
\end{equation}
so that, using the invariance of $\mu_H$,
we may write
\ba
\label{an eq}
R_{\g_s}({f}^0_{\g_s}) (A_{\g_s}) &=&
\int_{G^{k_{\g_s}}} [ f^0_{\g_s} \circ \psi^{-1}_{\g_s} \circ
\sigma_{\o{1},\g_s}] \nonumber \\
&& \prod_{i=1}^{N_{\g_s}}
\prod_{j=1}^{N_i+1} \rho\Big( g'_{(i,j)} [\prod_{k=j}^{N_i+1}
A(e_{(i,k)})]^{-1} \Big) d\mu_H(g'_{(i,j)})    \     .
\ea
{}From the analyticity of $\rho$
\cite{Ha} and the compactness of $G^{k_{\g_s}}$ it follows
that $R_{\g_s} (f^0_{\g_s})$ is a real-analytic function.
This concludes the proof of Lemma 5.

\end{subsection}

\begin{subsection} {Gauge covariance}

We now derive

\begin{lemma}       \label{ucl7}
$Z$ is a $\o\G$-covariant transform.
\end{lemma}

In particular, this will show that $R$ maps gauge
invariant functions to gauge invariant functions.

{\it Proof} \ For cylindrical $f = f_{\gamma} \circ
p_{\gamma}$,
\be
R_\g({f}_{\gamma}) (A_{\gamma}) =
\int_{G^{2E_\g}} (f_{\gamma} \circ \psi_{\gamma}^{-1})
(g_{1a} \o{A}(e_1)g_{1b}^{-1},..., g_{E_{\gamma}a}
\o{A}(e_{E_{\gamma}}) g_{E_{\gamma}b}^{-1})d\rho'_{\gamma}
\ee and
\ba
R_\g({f}_{\gamma}) (g_{\gamma}[A_{\gamma}]) = &&
\int_{G^{2E_\g}} (f_{\gamma} \circ \psi_{\gamma}^{-1})
(g_{1a} \o{g}_{p_{1a}} \o{A}(e_1) (\o{g}_{p_{1b}})^{-1}
 g_{1b}^{-1},..., \nonumber \\ && g_{E_{\gamma}a}
\o{g}_{p_{E_{\gamma} a}} \o{A}(e_{E_{\gamma}})
(\o{g}_{p_{E_{\gamma}b}})^{-1} g_{E_{\gamma}b}^{-1})
d\rho'_{\gamma}       \     ,
\ea
where $\o{g}_{p_{ia}}$ is the group element associated with the
initial vertex of edge $i$ by $g_\g \in {\cal G}_\g$
and $\o{g}_{p_{ib}}$ is the group element
associated with the final vertex of edge $i$.  Note that, in this
scheme, a point may be referred to as the initial and/or final vertex
of many edges.

We now perform the change of integration variables:
\ba
\label{change var} g_{ia} & \rightarrow & \o{g}_{p_{ia}}^{-1} g_{ia}
(\o{g}_{p_{ia}}) \nonumber \\
 g_{ib} & \rightarrow &
\o{g}_{p_{ib}}^{-1} g_{ib} (\o{g}_{p_{ib }})   \   .
\ea
The measure $d\rho'_{\gamma}$ contains only heat kernel-like
measures and delta
functions $\delta(g_{vi},g_{vj})$, where the notation indicates that
the arguments of a given delta-function are associated with the same
vertex $v$.  Since each such delta-function is unaffected by the above
transformation and the heat kernel-like functions $\rho_{\nu}$ are
conjugation invariant, $\rho'_\g$ is also invariant under (\ref{change
var}). Thus,
\ba
R_\g({f}_{\gamma}) (g_{\gamma}[A_{\gamma}]) &=&
\int_{G^{2E_\g}} (f_{\gamma}
\circ \psi_{\gamma}^{-1}) (\o{g}_{p_{1a}} g_{1a} \o{A}(e_1)
g_{1b}^{-1}\o{g}_{p_{1b}},..., \nonumber \\ && \
\o{g}_{p_{E_{\gamma} a}} g_{E_{\gamma}a}
\o{A}(e_{E_{\gamma}})
g_{E_{\gamma}b}^{-1}(\o{g}_{p_{E_{\gamma}b}})^{-1})
d\rho'_{\gamma}   \\
&=& R_\g({g_\g[f_\g]})(A_\g)  \nonumber
\ea
verifying gauge covariance for cylindrical $f$.  Since cylindrical
functions are dense in $L^2(\o{A},\tilde{\mu}_0)$, $L_{\o g}^*, L_{\o
g}^{\Co^*}$ are continuous $\forall \o g \in \o {\cal G}$ and we have
shown that $Z$ is an isometry and thus continuous, it
follows that $Z$ commutes with gauge transformations
and that Lemma \ref{ucl7} holds.  Theorem \ref{dccst} then follows as
a corollary of Lemmas \ref{ucl1}-\ref{ucl7}.

\bigskip

Before concluding, we note that a number of technical issues still
remain to be understood. Among these are the exact relationship of
$\agbc$ to $\abc / \gbc$ and a better understanding of the space
obtained by completing $L^2(\abc, \mu^{(\nu)}) \cap {\cal
H}_{\cal C}(\abc)$. It is also not known whether a {\it diffeomorphism
covariant} coherent state transform can be used to construct a
holomorphic representation from $L^2(\Ab, \mu_0)$. While we hope
that future investigation will clarify these matters, Theorems
\ref{th1} and \ref{dccst} as stated are enough to provide a framework
for the construction and analysis of holomorphic representations for
theories of connections.

\end{subsection}
\end{section}

\begin{section} {Acknowledgements}

We are  pleased to thank J. Baez, L. Barreira, A. Cruzeiro,
and C. Isham for useful discussions. Most of this research
was carried out at the Center for Gravitational Physics and
Geometry at The Pennsylvania State University and JL and JM would
like to thank the Center for its warm hospitality. The authors were
supported in part by the NSF Grant PHY93-96246 and the Eberly
research fund of The Pennsylvania State University.  DM was supported
in part also by the NSF Grant PHY90-08502. JL was supported in part
also by NSF Grant PHY91-07007, the Polish KBN Grant 2-P302 11207 and
by research funds provided by the Erwin Schr\"odinger Institute at
Vienna. JM was supported in part also by the NATO grant 9/C/93/PO and
by research funds provided by Junta Nacional de Investiga\c c\~ao
Cientifica e Tecnologica, STRDA/PRO/1032/93.

\end{section}

\begin{section}*{Appendix: The Abelian Case}

For compact Abelian $G$, the transform $Z$ of Section
\ref{DI} can be expressed in a particularly simple way and it is
possible to obtain explicit results.  We begin by simply evaluating
the transform of the holonomy $T_{\alpha}:\o{A} \rightarrow \Co^N$
associated with an arbitrary piecewise
analytic path $\alpha$.  (Note that the
results above for $\Co$-valued functions on $\o{\cal A}$ hold for
functions that take values in any Hilbert space.) This holonomy is
cylindrical over any graph $\gamma$ in which the path $\alpha$ may be
embedded and may be written as
\be
T_{\alpha}(\o{A}) = \prod_{i=1}^{E_{\gamma}}
[\o{A}(e_i)]^{m_i} \ ,
\ee
where the integer $m_i$ is the
(signed) number of times that the path $\alpha$ traces the
edge $e_i$.  Thus, the transform is given by the Baez integral over
$T_\alpha$
\ba    \label{0106}
R({T}_{\alpha})(\o{A}) &=& \int_{G^{E_{\gamma}}_a \times
G^{E_{\gamma}}_b} \prod_{i=1}^{E_{\gamma}}
[g_{ia}\o{A}(e_i)g_{ib}^{-1}]^{m_i} d\rho'_{\gamma}  \nonumber \\
 &=&
T_{\alpha}(\o{A}) \int_{G^{E_{\gamma}}_a \times
G^{E_{\gamma}}_b} \prod_{i=1}^{E_{\gamma}}
[g_{ia}g_{ib}^{-1}]^{m_i} d\rho'_{\gamma}
\ea
and $R$
is a scaling transformation on $T_{\alpha}$. Denote the resulting scaling
factor
for $T_\alpha$ on the right hand side of (105) by $e^{-l(\alpha)}$, that is,
$R[{T}_{\alpha}] = e^{-l(\alpha)} T_{\alpha}$. For
the case where $\nu$ is a Gaussian measure in standard
coordinates, we will show that $l(\alpha)$ is real and
positive.

Introduce coordinates $\theta \in [0,2\pi], r \in
(-\infty,\infty)$ on $U(1)^{\Co}$ such that $g^\Co = e^{i\theta}
e^r$.  We wish to consider a measure
$d\nu_{\sigma} = e^{-r^2/{\sigma}} {{d\theta dr} \over
{2\pi \sqrt{\pi \sigma}}}$  and the corresponding heat
kernel measure $d\rho_{\sigma}(\theta) = \sum_{k \in {\bf Z}}
e^{-[(\theta + 2\pi k)^2]/2{\sigma}}{{d\theta} \over {\sqrt{2
\pi \sigma}}}$. From (\ref{0106}) we find that
\be
e^{-l(\alpha)} = \int_{G^{k_{\gamma}}} \prod_{j=1}^{k_{\gamma}}
e^{iq_j\theta_j}d\rho_{\sigma}(\theta_j) = e^{-{{\sigma}
\over 2} \sqrt{{\sigma} \over {2 \pi}} \sum_jq^2_j}
\ee
for some $q^j \in {\bf Z}$ so that $l(\alpha)$ is real and positive,
as claimed.  Furthermore, since $q_j$ is a linear function of the
$m_i$, $l(\alpha) = \sum_{i,j} g^{ij}_{\gamma}m_im_j$ for some
symmetric matrix $g_{\gamma}^{ij}$ defined by $\gamma$,
${\psi_{\gamma}}$ and $\lambda_{\gamma}$.  The matrix
$g_{\gamma}^{ij}$ defines a Laplacian operator $\Delta_{\gamma} =
\sum_{i,j} g_\gamma^{ij} {{\partial} \over {\partial \theta_i}}
{{\partial} \over {\partial \theta_j}}$ on $G^{E_{\gamma}}$ and thus a
Laplacian on ${\cal A}_{\gamma}$, and our transform is the
corresponding coherent state transform on ${\cal A}_\g$.
Consistency of our transform
ensures that the $\Delta_{\gamma}$ are a consistent set of operators
and that they define a Laplacian $\Delta$ on some dense domain in
$L^2(\o{\cal A}, \mu_0)$.
Our transform is just the coherent state transform
on $\Ab$ defined by the heat kernel of the Laplacian
$\Delta$.

\end{section}

\end{document}